\documentclass[prd,aps,twocolumn,floats,floatfix,nofootinbib,superscriptaddress] {revtex4-1}
\usepackage{graphicx,subfigure}
\usepackage{epsfig}
\usepackage{mathrsfs}
\usepackage{amsmath}
\usepackage{amsfonts}
\usepackage{amssymb}
\usepackage{cancel}
\allowdisplaybreaks
\usepackage[usenames]{color}
\usepackage[a4paper,left=2.cm,right=2.cm,top=1.7cm,bottom=1.7cm]{geometry}
\usepackage{xcolor}
\usepackage{hyperref}
\usepackage{verbatim}
\usepackage[normalem]{ulem} 
\usepackage{placeins}

\hypersetup{
	colorlinks=true,
	linkcolor=red,
	linkbordercolor=black,
	citecolor=blue,
	filecolor=blue,      
	urlcolor=blue,
}

\begin{document}

\title{ Neutron star evolution with the Bemfica-Disconzi-Noronha-Kovtun viscous hydrodynamics framework}

\author{ Harry L. H. Shum}
\address{Nottingham Centre of Gravity, University of Nottingham, Nottingham NG7 2RD, UK}
\address{School of Mathematical Sciences, University of Nottingham, Nottingham, NG7 2RD, UK}
\author{Fernando Abalos}
\address{Departament de F\'isica, Universitat de les Illes Balears, Palma de Mallorca, E-07122, Spain }
\address{Institute of Applied Computing $\&$ Community Code (IAC3), Universitat de les Illes Balears, Palma de Mallorca, E-07122, Spain } 
\author{Yago Bea}
\address{Departament de Física Quàntica i Astrofísica and Institut de Ciències del Cosmos (ICC), Universitat de Barcelona, Martí i Franquès 1, ES-08028, Barcelona, Spain}
\author{Miguel Bezares}
\address{Nottingham Centre of Gravity, University of Nottingham, Nottingham NG7 2RD, UK}
\address{School of Mathematical Sciences, University of Nottingham, Nottingham, NG7 2RD, UK}
\author{Pau Figueras}
\address{School of Mathematical Sciences, Queen Mary University of London,
 Mile End Road, London E1 4NS, United Kingdom}
\author{Carlos Palenzuela}
\address{Departament de F\'isica, Universitat de les Illes Balears, Palma de Mallorca, E-07122, Spain }
\address{Institute of Applied Computing $\&$ Community Code (IAC3), Universitat de les Illes Balears, Palma de Mallorca, E-07122, Spain }

\begin{abstract}
The recently proposed first-order viscous relativistic hydrodynamics formulation by Bemfica, Disconzi, Noronha, and Kovtun (commonly known as the BDNK formulation) has been shown to be causal, stable, strongly hyperbolic, and thus locally well-posed. It is now a viable new option for modelling out-of-equilibrium effects in fluids, and has attracted wide attention in its potential applications to astrophysical systems.
In this work, we present the first non-linear numerical simulation of spherically symmetric neutron stars using the BDNK formulation under the Cowling approximation.  Using a simplified equation of state, we show that stable evolutions can be constructed within a restricted parameter space up to the simulation time we explored.  From these simulations, we analyse the frequency content of the quasi-normal modes and the decay rate of the fundamental mode.  This analysis serves as a first step towards constructing a fully consistent model of neutron stars using the BDNK formulation.
\end{abstract}

\maketitle

\tableofcontents

\section{Introduction}

Binary neutron star (NS) mergers are among the most fascinating astrophysical phenomena in the universe. They offer a unique laboratory for probing fundamental physics under extreme conditions. 
The observation of the event GW170817 \cite{Abbott:2017gw170817,Drout:2017ijr,Cowperthwaite:2017dyu} initiated the field of multi-messenger astronomy and has motivated further experimental and theoretical work. 
These systems provide a window into the properties of strongly interacting matter at high densities, the creation of heavy elements, the strong regime of gravity, and the nature of gravitational waves.
While current gravitational wave detectors primarily capture the inspiral phase of the NS merger, future observatories such as the Einstein Telescope \cite{Abac:2025saz} and Cosmic Explorer \cite{evans2021horizonstudycosmicexplorer} will access the post-merger regime, in which rich physical processes are present.

Numerical simulations of binary neutron star mergers play a central role in these studies. So far, the majority of such simulations treat the neutron star matter as an ideal fluid. The motivation behind this assumption is that the thermalization timescale  is set by the microscopic timescale of quantum chromodynamics (QCD), which is of the order of $10^{-23}$ seconds, while the typical macroscopic evolution timescale of the binary neutron star merger is of the order of $10^{-3}$ seconds.  Therefore, the two relevant timescales are separated by $20$ orders of magnitude and  the assumption that system is in local thermal equilibrium (and hence neglecting the viscous effects from QCD) should be a very good approximation.

To produce simulations which closely resemble astrophysical systems, all relevant physics must be included. In this context, it is important to analyse whether weak processes may play a significant role in the out-of-equilibrium dynamics of neutron stars.
Initial estimations of the possible relevance of weak processes suggested that the restoration of beta equilibrium during the highly dynamical post-merger regime might operate at similar timescales as the macroscopic dynamics of the system, 
thus giving rise to effective dissipative effects \cite{Alford:2017rxf}. 
These first estimations motivated further studies, and evaluating the consequences of the dissipative effects in neutron star mergers has become an active area of research. 

There has been recent progress in the microscopic studies of
weak processes in the background of neutron star matter, including hadronic \cite{Alford:2020lla,Alford:2021ogv,Alford:2022ufz} and quark matter \cite{Hernandez:2024rxi, CruzRojas:2024etx}, accounting for possible phase transitions to quark matter. 
Similarly, there have been efforts in quantifying the effects of weak processes in the numerical simulations of neutron star mergers \cite{Most:2021zvc,Most:2022yhe,Radice:2021jtw}. 
The picture that has emerged from these studies is that, while the QCD degrees of freedom are thermalized very fast, the weak degrees of freedom 
might not be at local thermal equilibrium.
Discerning whether they are near equilibrium or far from equilibrium is still an open question, and addressing it depends both on the microscopic computations and numerical simulations.  Assuming that the system is near thermal equilibrium, a description of the dynamics of the system based on viscous hydrodynamics should be adequate. 
Therefore, carrying out simulations of neutron star mergers with viscous effects has become a pressing challenge.
A groundbreaking result on this front was achieved by Chabanov and Rezzolla \cite{Chabanov:2021dee,Chabanov:2023abq,Chabanov:2023blf}, who were the first ones to simulate neutron star binary mergers including viscosity. 
However, further studies are needed to understand the impact of viscous effects on the various observables.
Other numerical works implementing viscous effects include studies in spherical symmetry \cite{Camelio:2022fds} and in related scenarios \cite{Radice:2017zta,Shibata:2017jyf,Shibata:2017xht,Duez:2020lgq}.

Introducing viscosity in a theory of relativistic hydrodynamics  presents some difficulties that we now describe. As originally formulated decades ago, the relativistic versions of the Navier-Stokes equations written by Eckart \cite{Eckart:1940te} and Landau and Lifshitz \cite{LandauBook} respectively, were found to be mathematically ill-posed \cite{Hiscock:1985zz}, and hence inadequate to describe the dynamics of real-world relativistic viscous fluids. 
An alternative formulation, which is mathematically well-posed (at least in some regimes),  was put forward by Müller, and Israel and Stewart  (MIS) \cite{Muller:1967zza,Israel:1976tn,Israel:1979wp}. In this MIS approach, the hydrodynamic expansion is extended up to second order in derivatives of the thermodynamic variables, and the dissipative part of the stress tensor is promoted to a new variable, with its own  \textit{ad-hoc} evolution equation. By introducing the new variables and their corresponding evolution equations, the new set of evolution equations enjoys better mathematical  properties. 
Clearly this procedure is not unique, and hence one does not refer to a single MIS formulation of relativistic viscous hydrodynamics.
More recently, the MIS approach has been applied to construct new theories of viscous hydrodynamics (see e.g., \cite{Baier:2007ix,Denicol:2012cn}), some of which are implemented in modern codes that have been successfully used to describe the experimental data of the quark-gluon plasma produced in heavy-ion collisions. In these cases, viscous effects have been found to be important to correctly interpret the data.
Other mathematically consistent theories of dissipative fluids were put forward by Geroch and Lindblom \cite{Geroch:1990bw}, i.e., divergence-type theories. Remarkably, it was later shown that, under suitable conditions and assumptions, MIS theories and divergence-type theories have the same physical content \cite{Geroch:1995bx,Lindblom:1995gp}.

Despite their successes, MIS formulations present some important limitations. It was not until recently that the first mathematical proof of well-posedness for a specific MIS formulation was obtained \cite{Bemfica:2020xym}, while for other MIS formulations this analysis is still lacking.
Furthermore, by checking the well-posedness conditions of the MIS equations in realistic hydrodynamic simulations of the quark-gluon plasma, unsurprisingly it was found that there can be significant violations, mostly at the initial stages where the viscous terms are not small \cite{Plumberg:2021bme,Chiu:2021muk,ExTrEMe:2023nhy,Domingues:2024pom}. These results indicate that the theory is being applied outside its regime of validity and hence it might not provide an accurate description of the experimental data. 
Therefore, to model the experimental data, one needs to use a theory that at least has a well-posed initial value problem,  like the theory that we now present.

Recently Bemfica, Disconzi and Noronha \cite{Bemfica:2017wps,Bemfica:2019knx} and Kovtun \cite{Kovtun:2019hdm} (BDNK), proposed a new well-posed formulation of first order relativistic viscous hydrodynamics based on an effective field theory approach. In this new formulation, one considers arbitrary field redefinitions of the thermodynamic variables up to first order in derivatives, which defines an arbitrary hydrodynamic frame (i.e., an extension of the thermodynamic variables off equilibrium). By imposing causality and well-posedness, they identified open sets of hydrodynamic frames for which the theory is mathematically sound. Unsurprisingly, the Landau and Eckart frames are excluded from the allowed frames.

These mathematical results were followed by some numerical works that tested this new formulation in highly non-trivial settings \cite{Pandya:2021ief,Bantilan:2022ech,Bea:2023rru,Pandya:2022pif,Pandya:2022sff,keeble2025firstorderviscousrelativistichydrodynamics,Hatton:2025crr,fantini2025constraintinstabilitiesfirstorderviscous,Hatton:2024jyt}. 
In particular, \cite{Bea:2025eov} used the BDNK theory  to describe experimental data of the quark-gluon plasma created in heavy-ion collisions at the LHC, since it offers potential solutions to some of the problems present in the state-of-the-art codes based on the MIS formulation. These successes further motivate the application of the BDNK formulation of relativistic viscous hydrodynamics to astrophysical systems.

The MIS formulations seem to be a natural framework to describe viscous effects in neutron mergers since, until recently, this has been the standard approach for describing viscous relativistic fluids. 
In fact, the recent work of Chabanov and Rezzolla \cite{Chabanov:2021dee,Chabanov:2023abq,Chabanov:2023blf} uses a particular MIS formulation. Thus, one can check if the problems that MIS formulations present in other contexts, can also arise in realistic numerical simulatoins of binary neutron star mergers. Interestingly, in these papers the authors checked the well-posedness conditions obtained in \cite{Bemfica:2020xym}, and they find regions of spacetime where these conditions are not satisfied. 

On the other hand, at the heart of the BDNK framework is the freedom to choose the hydrodynamic frame, an important property that MIS does not have.\footnote{Typically, MIS formulations work in the Landau or Eckart frames, since they are very natural in terms of organising the hydrodynamic expansion.} Using this freedom, one can choose the parameters/functions that specify the frame to be suitable functions of the system's variables and thus prescribe the characteristic velocities of the system. These velocities can be specified a priori, independently of initial data or the evolution of the variables (and their derivatives), and hence ensure the well-posedness of the evolution equations. Therefore, the BDNK framework can potentially be  well-posed in a wider range of situations than the corresponding MIS formulations, which provides strong motivation to explore viscous effects in neutron star mergers within this framework. That said, a word of caution is due; even if the BDNK evolution equations are well-posed, the system may not be in the hydrodynamic regime, in which case the BDNK framework would not necessarily provide an accurate description of the relevant physics.

In this paper, we go beyond the state-of-the-art and use the BDNK equations to simulate the dynamics of neutron stars.\footnote{We note that recently,  linear perturbations of viscous neutron stars have been studied using the BDNK formulation \cite{caballero2025neutronstarradialperturbations,Redondo-Yuste:2025ktt} and other viscous relativistic hydrodynamics frameworks \cite{Mendes:2025oib}.} More precisely, we study radial, spherically symmetric oscillations of an isolated neutron star treating the fluid fully non-linearly whilst keeping gravity non-dynamical for simplicity (Cowling approximation). We use a simple model for the equation of state and transport parameters as a first study in this direction. Moreover, we do not include the evolution of the rest mass density equation in our description. In this setup, we perform long-time evolutions and study the non-linear evolution of small perturbations, as well as evaluate the effect of the viscosity and the use of different hydrodynamic frames.

This paper is structured as follows.  In Section \ref{section:BDNK_formulation}, we provide a brief overview of the BDNK formulation.  We will outline its 3+1 decomposition and the adaptations to spherical symmetry, including the numerical methods we use for performing simulations.  The equation of state is also specified.  In Section \ref{simuNS}, we present the numerical results of our simulations.  This is followed by an analysis of the quasi-normal mode frequency spectrum and the decay rate of the fundamental mode.
Throughout this paper, we work in geometric units $G=c=1$.  Unless otherwise stated, we use Greek letters $\mu, \nu, ...$ that run from $0$ to $3$ to denote spacetime indices, and Roman indices $i,j,...$ that run from $1$ to $3$ to denote spatial indices.  We adopt Einstein's convention for summing over indices, and our metric signature is taken to be mostly positive.

\section{BDNK formulation of relativistic viscous hydrodynamics}\label{section:BDNK_formulation}

Consider the stress-energy tensor of an ideal (non-conformal) fluid: 
\begin{align}
T^{\mu\nu} = \epsilon \,u^\mu\,u^\nu + p(\epsilon)\,\Delta^{\mu\nu}\,,
\label{eq:stress_tensor0}
\end{align}
where $\epsilon$ represents the total energy density of the fluid, $u^\mu$ is the four-velocity, and $p$ is the pressure, which is determined by the equation of state (EoS) $p=p(\epsilon)$.
We have also introduced the projector onto the space orthogonal to $u^\mu$, namely
\begin{align}
\Delta^{\mu\nu}=g^{\mu\nu}+u^\mu\,u^\nu\,.
\end{align}

The equations of motion follow from projecting the conservation of the stress-energy tensor
\begin{align}
\nabla_\mu T^{\mu\nu}=0 \label{eq:conservation}
\end{align}
in the direction along and orthogonal to $u^\mu$. By using the stress-energy tensor in Eq.(\ref{eq:stress_tensor0}), these projections give
\begin{align}
&u^\mu\nabla_\mu \epsilon+(\epsilon+p)(\nabla_\mu u^\mu)=0\,, \label{eq:ideal_eom_non_conformal1}\\
&u^\nu\nabla_\nu u^\mu+\frac{p'(\epsilon)}{\epsilon+p}\,\Delta^{\mu\nu}\nabla_\nu\epsilon=0\,. \label{eq:ideal_eom_non_conformal2}
\end{align}

Hydrodynamics can be thought of as the effective field theory (EFT) describing the dynamics of fluids near thermodynamic equilibrium. In this context, the ideal fluid stress-energy tensor \eqref{eq:stress_tensor0} is the leading order term in a derivative expansion of infrared (IR) variables,  namely the energy density and velocity of the fluid in this particular case. In the ideal fluid stress-energy tensor \eqref{eq:stress_tensor0}, these variables appear with no derivatives. Dissipative effects are incorporated as new terms in the stress-energy tensor of the fluid with non-vanishing derivatives of the thermodynamic fields. It is well-known that this derivative expansion is non-unique because the thermodynamic variables are not uniquely defined away from equilibrium (see e.g., Ref.~\cite{Kovtun:2019hdm}). Every choice of the thermodynamic variables away from equilibrium corresponds to a specific choice of hydrodynamic frame. For instance, the widely used Landau frame is defined such that $u_\mu\,T^{\mu\nu}_\text{dissipative}=0$, and it is well-known that the initial value problem for relativistic dissipative fluids in this frame is ill-posed \cite{Hiscock:1985zz}.\footnote{The same applies to Eckart's frame \cite{Eckart:1940te}, but this is not relevant for us since we are not considering charge density.}

In the context of EFTs, it is natural to consider the most general stress-energy tensor at every order in the derivative expansion. This was the basic idea of \cite{Bemfica:2017wps,Kovtun:2019hdm}, who considered field redefinitions of the thermodynamic variables up to first order in derivatives, and showed that there exist choices of hydrodynamic frames for which the initial value problem is well-posed. Unsurprisingly, such choices exclude the Landau frame. The resulting formulation of viscous relativistic fluids is known as BDNK. In this paper, we will follow BDNK and consider the following form of the stress-energy tensor for relativistic viscous fluids up to first order in derivatives:
\begin{align}
T^{\mu\nu}  = &~ (\epsilon + \mathcal{A})\,u^\mu\,u^\nu+(p+\Pi)\,\Delta^{\mu\nu}+\mathcal{Q}^\mu\,u^\nu\nonumber \\
   & \quad +~u^\mu\,\mathcal{Q}^\nu-2\,\eta\,\sigma^{\mu\nu}\,,    
\label{eq:vis_stress_tensor}
\end{align}
where the new terms in the stress-tensor are given by
\begin{align*}
\mathcal{A} &= \tau_\epsilon\left[ u^\mu\nabla_\mu\epsilon + (\epsilon+p)(\nabla_\mu u^\mu)\right]\,,\\
\Pi &= -\zeta\,\nabla_\mu u^\mu + \tau_p\left[ u^\mu\nabla_\mu\epsilon + (\epsilon+p)(\nabla_\mu u^\mu)\right]\,, \\
\mathcal{Q}^\mu &= \tau_Q(\epsilon + p)\,u^\nu\nabla_\nu u^\mu+\beta_\epsilon\,\Delta^{\mu\nu}\nabla_\nu\epsilon\,,    
\end{align*}

and
\begin{align}
    \sigma^{\mu\nu} = \tfrac{1}{2}\left[\Delta^{\mu\alpha}\Delta^{\nu\beta}(\nabla_\alpha u_\beta+\nabla_\beta u_\alpha)\right.\nonumber \\
    \left.\qquad     -\tfrac{2}{3}\,\Delta^{\mu\nu}\Delta^{\alpha\beta}\nabla_\alpha u_\beta\right]\, \label{eq:shear}
\end{align}
is the shear tensor. 
Note that the new terms in \eqref{eq:vis_stress_tensor} contain first derivatives of the energy density and the four-velocity of the fluid. The viscous stress-energy tensor \eqref{eq:vis_stress_tensor} is not the most general one that arises from the field redefinitions considered in \cite{Bemfica:2017wps,Kovtun:2019hdm,Hoult:2020eho}. Here, we have already made the same choices of the coefficients in the derivative expansion as in \cite{Bemfica:2020zjp} to guarantee that the corresponding theory of relativistic viscous fluids is i) causal, ii) stable, and iii) strongly hyperbolic (hence locally well-posed), see Ref.~\cite{Bemfica:2020zjp} for the proofs.

The terms $\tau_\epsilon$, $\zeta$, $\tau_p$, $\tau_Q$, $\beta_\epsilon$, $\eta$ in \eqref{eq:vis_stress_tensor} are known as transport coefficients and, in our particular case, they are functions of the energy density $\epsilon$. The coefficients $\zeta$ and $\eta$ determine the strength of the dissipative effects and they are the only transport coefficients that are independent of the choice of hydrodynamic frame at this order in the derivative expansion. In principle, one can compute them given a microscopic theory.  This is the reason why these coefficients are regarded as the physical transport coefficients and they are given special names, namely the bulk viscosity and the shear viscosity, respectively. The coefficients $\tau_{\epsilon}$, $\tau_{p}$ and $\tau_{Q}$ are known as the relaxation times, as they determine the time scales over which the dissipation occurs. Finally, the coefficient $\beta_{\epsilon}$ controls the contribution of the energy density to the heat flux.  The choice of transport coefficients corresponds to specifying a certain hydrodynamic frame. It is worth noticing that, for our choice of the viscous stress-energy tensor \eqref{eq:vis_stress_tensor}, some of the viscous contributions included within $\mathcal{A}$ and $\Pi$ (but not $\mathcal{Q}^{\mu}$) contain terms that are proportional to the ideal equations of motion. We will come back to this point below.

At this stage, we make yet another special choice of field redefinitions at this order in derivatives.  To ensure that $\mathcal{Q}^{\mu}$ is  proportional to the zeroth order equation of motion \eqref{eq:ideal_eom_non_conformal2}, we choose $\beta_\epsilon = \tau_Q\,p'(\epsilon)$,\footnote{We note that \cite{Pandya:2022sff} made a similar choice in their more general setting.}  so that 
\begin{align}
\mathcal{Q}^\mu = \tau_Q\left[(\epsilon + p)\,u^\nu\nabla_\nu u^\mu+p'(\epsilon)\,\Delta^{\mu\nu}\nabla_\nu\epsilon\right]\,.
\end{align}

Making field redefinitions that are proportional to the lower-order equations of motion is the usual procedure in classical EFTs; it effectively reduces the freedom in making such field redefinitions and it ensures that the contributions of the terms in the expansion that are sensitive to the choice of hydrodynamic frame are subdominant with respect to those terms that are invariant at the same order in the derivative expansion. This is the reason why the latter are considered to be  ``physical''. In the case of relativistic viscous hydrodynamics, the particular choice of frame that we have made ensures that the dissipative effects encoded in the shear and bulk viscosity terms dominate over the remaining first derivative terms. In fact, with our choice and in the regime of validity of hydrodynamics, the latter are of higher order precisely because they are proportional to the ideal equations of motion. A more detailed discussion of the effects of the choice of hydrodynamic frame in conformal relativistic viscous hydrodynamics can be found in \cite{Bea:2023rru}.\\

\subsection{3+1 decomposition of BDNK equations}

The conservation equation \eqref{eq:conservation} can be written as an evolution system by performing the standard 3+1 decomposition. First, the line element is explicitly decomposed into time and coordinate components, namely
\begin{align}
ds^2 &= g_{\mu \nu} dx^{\mu} dx^{\nu} \label{metric}\\
&= -\alpha^2\,dt^2+\gamma_{ij}(dx^i+\beta^i\,dt)(dx^j+\beta^j\,dt)\,.\nonumber 
\end{align}
where $\alpha$ is the lapse function, $\beta^i$ is the shift vector and $\gamma_{ij}$ is the induced metric on the spatial hypersurfaces defined by $t=\text{const}.$. For a more geometrical formulation, one can define $n_\mu=-\alpha\,(dt)_\mu$ as the unit timelike co-vector orthogonal to these hypersurfaces, with $n^\mu=\frac{1}{\alpha}(\partial_t^\mu-\beta^i\partial_i^\mu)$ the associated timelike unit vector, i.e., $n_{\mu} n^{\mu}=-1$. 
Therefore, the induced metric can be written as
\begin{align}
\gamma_{\mu\nu} = g_{\mu\nu} +n_\mu\,n_\nu\,,    
\end{align}
while we can introduce the extrinsic curvature as
\begin{align}
K_{\mu\nu} &=-\tfrac{1}{2}\,\mathcal{L}_n\gamma_{\mu\nu} \\
&=~-(\nabla_\mu n_\nu+n_\mu\, a_\nu)\,,
\end{align}
where $a_\mu = n^\nu\nabla_\nu n_\mu$ is known as the acceleration of the normal vector. Note that $a^\mu$ is a spatial vector, i.e.,  $n^\mu a_\mu=0$. 

After these preliminaries, we can perform the same decomposition for the fluid fields. The four-velocity  $u^\mu$ can be decomposed in terms of $n^\mu$ and the spatial velocity $v^\mu$ (i.e., $n^\mu v_\mu=0$), namely
\begin{align}
u^\mu = W(n^\mu + v^\mu)\,,
\label{eq:def_u_A}
\end{align}
where
\begin{align}
W=\frac{1}{\sqrt{1-v_\mu v^\mu}} = \frac{1}{\sqrt{1-\gamma_{ij}v^i v^j}} \label{eq:def_W}\,,
\end{align}
is the usual Lorentz factor. The components of the velocity vector $v^\mu$ can be written explicitly 
\begin{align}
v^t=0\,,\quad v^i=\tfrac{1}{\alpha}\left(\tfrac{u^i}{u^t}+\beta^i\right)\,.
\end{align}

Similarly, the decomposition of a general stress-tensor $T^{\mu\nu}$ in components tangent and perpendicular to the spatial hypersurfaces is given by, 
\begin{align}
T^{\mu\nu} = E\,n^\mu\,n^\nu + S^\mu\,n^\nu + n^\mu\,S^\nu + S^{\mu\nu}\,,
\end{align}
with
\begin{align}
E &=n_\mu\,n_\nu\,T^{\mu\nu}\,,\quad S^\mu = -\gamma^\mu_{\phantom{\mu}\alpha}\,n_\beta\,T^{\alpha\beta}\,,\nonumber\\
 S^{\mu\nu} &= \gamma^\mu_{\phantom{\mu}\alpha}\gamma^\nu_{\phantom{\nu}\beta}\,T^{\alpha\beta}\,.
\label{eq:projections}
\end{align}
These projections correspond to the energy density, the linear momentum density and the tensor of stresses, respectively.
Again, in the coordinates \eqref{metric},  the different components of the stress tensor are,
\begin{align}
T^t_{\phantom{t}t} &= -E\,,\quad T^t_{\phantom{t}i} = \tfrac{1}{\alpha}\,S_i\,,\nonumber\\ T^i_{\phantom{i}t} &=\beta^i-\alpha\,S^i\,,\quad T^i_{\phantom{i}j} = -\tfrac{1}{\alpha}\,\beta^i\,S_j+S^i_{\phantom{i}j}\,.
\end{align}

The projection of the conservation equation \eqref{eq:conservation} gives a system of balance laws,
\begin{align}
    \partial_t\mathbf{q} +\partial_i\mathbf{F}^i(\mathbf{q})=\mathbf{S}(\mathbf{q})\,. \label{eq:balance_law}
\end{align}
In terms of the projections \eqref{eq:projections}, they are given by
\begin{widetext}
\begin{align}
\partial_t(\sqrt{\gamma}\,E)+\partial_i[\sqrt{\gamma}(\alpha\,S^i - \beta^i\,E)]&=\alpha\sqrt{\gamma}(S^{ij}\,K_{ij} - S^i\partial_i\ln\alpha)\,,\label{eq:eq_E}\\
\partial_t(\sqrt{\gamma}\,S_j)+\partial_i[\sqrt{\gamma}(\alpha\,S^i_{\phantom{i}j}-\beta^i\,S_j)]&=\alpha\sqrt{\gamma}\left(\tfrac{1}{2}\,S^{ik}\partial_j\gamma_{ik}+\tfrac{1}{\alpha}\,S_i\partial_j \beta^i-E\,\partial_j\ln\alpha\right)\,, \label{eq:eq_Si}
\end{align}
\end{widetext}
from which we identify the conservative variables
\begin{align}
    \mathbf{q}=(\sqrt{\gamma}\,E,\,\sqrt{\gamma}\,S_i)\,,
\end{align}
the fluxes
\begin{align*}
    \mathbf{F}=\big(\sqrt{\gamma}(\alpha\,S^i - \beta^i\,E),\,\sqrt{\gamma}(\alpha\,S^i_{\phantom{i}j}-\beta^i\,S_j)\big)\,,
\end{align*}
and the sources
\begin{widetext}
\begin{align*}
    \mathbf{S} &= \left(\alpha\sqrt{\gamma}(S^{ij}\,K_{ij} - S^i\partial_i\ln\alpha),\alpha\sqrt{\gamma}\left(\tfrac{1}{2}\,S^{ik}\partial_j\gamma_{ik}+\tfrac{1}{\alpha}\,S_i\partial_j \beta^i-E\,\partial_j\ln\alpha\right)\right)\,.
\end{align*}    
\end{widetext}

Notice that both the fluxes and the sources cannot be explicitly written in terms of the conserved quantities alone. Therefore, the numerical evolution requires a preliminary mapping from the conservative variables $\mathbf{q}$ into the primitive variables $\mathbf{p}=(\epsilon,v^{i})$ by solving the constitutive equations, which also involve the equation of state $p=p(\epsilon)$. This recovery of the primitive fields is a well-known procedure in ideal fluids. However, in the viscous case, there is a subtle but important difference: the stress-energy tensor \eqref{eq:vis_stress_tensor} contains first derivatives (both spatial and temporal) of the thermodynamic variables. 
To proceed, we perform the first-order reduction in time. First, for the energy density, we consider\footnote{Notice that this choice differs from that in \cite{Pandya:2022pif}; ours seems more natural and easier to implement using standard High-Resolution Shock-Capturing methods.}
\begin{align}
\hat\epsilon &= -n^\mu\nabla_\mu\epsilon\,. \label{eq:eq_epsilon}
\end{align}
On the other hand, for the spatial velocity, one would naively consider
\begin{align}
\hat v^{\mu} &=-n^\nu\nabla_\nu v^\mu\,, \label{eq:eq_hat_v}
\end{align}
but $n_\mu\,\hat v^\mu = v^\mu\,a_\mu$, which is non-zero in general. From now on, in order to alleviate the lengthy expressions, we use the bar $\bar\,\,$ to indicate that the contraction of that tensor with the normal vector $n^\mu$ vanishes, so $a_\mu=\bar a_\mu$ and $v^{\mu}=\bar v^{\mu}$.  Therefore, we shall perform a first-order reduction of the spatial velocity that is also spatial by defining
\begin{align}
        \hat{\bar v}^\mu=~\gamma^\mu_{\phantom{\mu}\alpha} \hat v^\alpha 
=-\gamma^\mu_{\phantom{\mu}\alpha}n^\nu\nabla_\nu v^\alpha\,.
    \label{eq:eq_hatbv}
\end{align}
Then, equations \eqref{eq:eq_epsilon} and \eqref{eq:eq_hatbv} provide the desired first order reduction in time, and when written in 3+1 form, they are the evolution equations for the total energy density $\epsilon$ and spatial velocity $v^i$ respectively:
\begin{align}
(\partial_t-\beta^j\partial_j)\epsilon &=-\alpha\,\hat\epsilon\,, \label{eq:evolution_epsilon}\\
(\partial_t-\beta^j\partial_j)v^i &=\alpha(-\hat{\bar v}^i+K^i_{\phantom{i}j}\,v^j)-v^j\partial_j\beta^i\,. \label{eq:evolution_v}
\end{align}

At this point we can employ $\mathbf{p}_0=(\epsilon,\,v^i)$ and $\mathbf{p}_1=(\hat\epsilon, \, \hat{\bar v}^i)$ as our primitive variables.  By considering the projections \eqref{eq:projections} for \eqref{eq:vis_stress_tensor} it is straightforward to find the map between the primitive variables $\{\mathbf{p}_0,\,\mathbf{p}_1\}$ and the conservative variables $\{\mathbf{q}\}$. Since the evolution of $\mathbf{p}_0$ only depends on $\mathbf{p}_1$, we  only need to find $\mathbf{p_1}$ in terms of $\mathbf{q}$ and $\mathbf{p}_0$ (and their spatial derivatives). Notice that this mapping $\mathbf{p_1}= \mathbf{p_1} (\mathbf{q},\mathbf{p}_0)$ is linear because  \eqref{eq:vis_stress_tensor} is linear in the time derivatives of the thermodynamic variables.

Let us write the explicit relations between the conservative variables and the primitive ones. For the viscous stress tensor \eqref{eq:vis_stress_tensor}, we have:
\begin{widetext}
\begin{align}
E=&-(p+\Pi)+(\epsilon+\mathcal{A}+p+\Pi)W^2-2\,n_\alpha\,\mathcal{Q^\alpha}\,W-2\,\eta \,(n_\alpha \, n_\beta \, \sigma^{\alpha\beta})\,,\label{eq:E_fluid}\\
S^\mu=&~v^\mu(\epsilon+\mathcal{A}+p+\Pi)\,W^2+  (\gamma^\mu_{\phantom{\mu}\alpha}\,\mathcal{Q}^\alpha -n_\alpha \mathcal{Q}^\alpha v^\mu)W + 2\,\eta\,(\gamma^\mu_{\phantom{\mu}\alpha}\,n_\beta\,\sigma^{\alpha\beta})\,, \label{eq:Si_fluid}\\
S^{\mu\nu}=&~v^\mu\,v^\nu(\epsilon+\mathcal{A}+p+\Pi)\,W^2 + \mathcal{Q}^\alpha(v^\mu\,\gamma^\nu_{\phantom{\nu}\alpha}+v^\nu\,\gamma^\mu_{\phantom{\nu}\alpha})W
+(p+\Pi)\gamma^{\mu\nu} - 2\,\eta\,(\gamma^\mu_{\phantom{\mu}\alpha}\,\gamma^\nu_{\phantom{\nu}\beta}\,\sigma^{\alpha\beta})\,.\label{eq:Sij_fluid}
\end{align}
\end{widetext}

For the following calculations, it is useful to consider the following decomposition of the 3-velocity:
\vspace{-5pt}
\begin{align}
\nabla_\mu v_\nu = D_\mu  v_\nu + n_\mu\,\hat{\bar v}_\nu - v^\alpha\,K_{\alpha\mu}\,n_\nu- n_\mu\,n_\nu(v^\alpha a_\alpha)\,,
\end{align}
which results in the following projections: 
\begin{align}
\gamma_i^{\phantom{i}\mu}\gamma_{j}^{\phantom{j}\nu}\nabla_\mu v_\nu&= D_i v_j\,,\\
\gamma_i^{\phantom{i}\mu} n^\nu \nabla_\mu v_\nu&= K_{ij}\,v^j\,,\\
n^\mu\gamma_i^{\phantom{i}\nu}\nabla_\mu v_\nu &= -\hat{\bar v}_i\,,\\
n^\mu n^\nu \nabla_\mu v_\nu &=- v^i\,a_i = -v^i \partial_i\ln\alpha\,.
\end{align}
Then, the 3+1 decomposition of the stress-energy tensor can be written as

\begin{widetext}
\begin{align}
    E=&~W^2 \epsilon - p\, (1 - W^2) \nonumber\\
    &+ W\big[\tau_p (1 - W^2)-W^2\tau_\epsilon\big] \Bigl\{\hat\epsilon   -  v^{i} D_{i}\epsilon -  (\epsilon + p) \bigl[-K+v_{i} (a^{i} -  \hat{\bar v}^{i} W^2) + D_{i}v^{i} + W^2 v^{i} v^{j}  D_{i}v_{j}\bigr]\Bigr\} \nonumber\\
    &+2\,\tau_Q\,W\Bigl\{(1 - W^2)p'(\epsilon)\,\hat\epsilon + W^2 \bigl[p'(\epsilon)\, v^{i} D_{i}\epsilon +(\epsilon + p) \bigl(v^{i} v^{j}(- K_{ij} + W^2 D_{i}v_{j}) + v_{i} (a^{i} -  \hat{\bar v}^{i} W^2)\bigr) \bigr]\Bigr\} \nonumber \\ 
    & + \tfrac{2}{3}\,\eta\, W \Big\{ (1 - W^2) \bigl(K + 2 v_{i} (a^{i} -  \hat{\bar v}^{i} W^2) - D_{i}v^{i}\bigr) \nonumber\\
    &\hspace{1.5cm}+ W^2 v^{i} v^{j}  \bigl[3 K_{ij} + (1 + 2 W^2) \bigl((-2 + W^2) D_{i}v_{j} - W^2\,v_{i} v^{l}  D_{l}v_{j}\bigr)\bigr]
    \Big\}\nonumber\\
    &+\zeta\,W (1 - W^2) \Bigl[-K + v_{i} ( a^{i} - \hat{\bar v}^{i} W^2) +  D_{i}v^{i} + W^2 v^{i} v^{j}  D_{i}v_{j}\Bigr]\,, \label{eq:E_fluid_3p1}\\
    S^i=&~ v^{i} \,W^2 (\epsilon+p) \nonumber\\
    &+(\tau_p+\tau_\epsilon)\,v^{i} \,W^3 \Bigl\{-\hat\epsilon +  (\epsilon + p)\bigl[-K + \bigl(v_{j} (a^{j} - W^2\, \hat{\bar v}^{j} ) + D_{j}v^{j} + v^{j} v^{k} W^2 D_{j}v_{k}\bigr)\bigr] +  v^{j} D_{j}\epsilon\Bigr\}\nonumber\\
    &+\tau_Q\,W \Bigl\{W^2 (\epsilon + p) \bigl(a^{i} - \hat{\bar v}^{i} + v^{j} (-K^{i}{}_{j} +  D_{j}v^{i})\bigr) + v^{i} \Bigl[p'(\epsilon)\bigl( (1- 2\, W^2)\hat\epsilon  +2\,W^2\,v^{j} D_{j}\epsilon\bigr) \nonumber\\
    & + W^2 (\epsilon + p)\bigl(a^{j} v_{j}  - K_{jl}\, v^{j}\, v^{l} - 2\, W^2  (\hat{\bar v}^{j} v_{j} -  v^{j} v^{k} D_{j}v_{k})\bigr)\Bigr] +  p'(\epsilon)\, D^{i}\epsilon\Bigr\} \nonumber\\
    &+\eta\,W\Bigl\{ (1 - W^2)(a^{i}-\hat{\bar v}^{i}) +  K^{i}{}_{j}\, v^{j} (1 + W^2) - \tfrac{1}{3} W^2\bigl[
     v^i\bigl(
        2\, K + a^{j}\, v_{j} + 3\, \hat{\bar v}^{j}\, v_{j} - 3 \,K_{jl} \, v^{j} \,v^{l} - 2 \, D_{j}v^{j}\nonumber\\
        &\hspace{1.25cm}- 4\,W^2( \hat{\bar v}^{j} v_{j} -  v^{j} v^{l}  D_{j}v_{l})\bigr) +3(v^{j} D^{i}v_{j}+v^{j} D_{j}v^{i})
     \bigr]
    \Bigr\} \nonumber\\
    &-\zeta\,v^{i} W^3 \bigl[ -K + v_{j} (a^{j} -  \hat{\bar v}^{j} W^2) + D_{j}v^{j} + W^2\,v^{j} \,v^{l}\,  D_{j}v_{l}\bigr]\,, \label{eq:Si_fluid_3p1}\\
    S_{ij} =&~ p\,\gamma_{ij}+  W^2 (\epsilon+p)v_{i}\,v_{j} \nonumber\\
    &- W [\tau_p\,\gamma_{ij} + (\tau_\epsilon+\tau_p)W^2 v_{i}\, v_{j}]\! \cdot\!\Bigl[\hat\epsilon -  v^{l} D_{l}\epsilon+ K (\epsilon + p)  
    -  (\epsilon + p) \bigl(v_{l} (a^{l} - W^2 \hat{\bar v}^{l}) + D_{l}v^{l} +W^2 v^{m} v^{n} D_{m}v_{n}\bigr)\Bigr]\nonumber\\
    &+\tau_Q\Bigl\{
    2\,W\,v_{(i}\bigl[W^2 (\epsilon + p) \bigl(a_{j)} -  \hat{\bar v}_{j)} + v^{l} (- K_{j)l} + D_{|l|}v_{j)})\bigr) + p'(\epsilon) D_{j)}\epsilon\bigr] \nonumber\\
    &\hspace{1cm}+2\,W\,v_i\,v_j\bigl(- p'(\epsilon) \hat\epsilon - W^2 (\epsilon + p)( \hat{\bar v}^{l} v_{l}  - v^{l} v^{m}  D_{l}v_{m}) + p'(\epsilon)\, v^{l} D_l\epsilon\bigr)
    \Bigr\}\nonumber\\
    &+\tfrac{1}{3}\,\eta\, W \biggl\{
    6 \,K_{ij}+6\, W^2 \,v^{l}K_{l(i} v_{j)} - 2 \,W^2 (\gamma_{ij}-2\,W^2\,v_i\,v_j)\big(\hat{\bar v}^{l} v_{l}-v^m\,v^nD_m v_n\big)\nonumber\\
    &\hspace{1.5cm}+2\,(\gamma_{ij} +W^2\, v_{i} \,v_{j} )\big(-K+a^l\,v_l+D_lv^l\bigr)\nonumber\\
    &\hspace{1.5cm}-6\,\bigl[ D_{(i}v_{j)} 
    + W^2\bigl(  v^{l} v_{(i} D_{j)}v_{l} + v^{l}v_{(i}  D_{|l|}v_{j)} + (a_{(i} -  \hat{\bar v}_{(i}) v_{j)}\bigr)
    \bigr]
    \biggr\} \nonumber\\
    &+\zeta\,W (\gamma_{ij} +W^2\, v_{i}\, v_{j} ) \bigl[K + v_{l} (- a^{l} + W^2\hat{\bar v}^{l} ) -  D_{l}v^{l} - W^2 v^{m} v^{n}  D_{m}v_{n}\bigr]\,. \label{eq:Sij_fluid_3p1}
\end{align}
\end{widetext}

Our evolution strategy can then be summarised as follows. We have two sets of primitive variables, namely $\mathbf{p}_0=(\epsilon,\,v^i)$ and $\mathbf{p}_1=(\hat\epsilon,\,\hat{\bar v}^i)$,  conservative variables $\mathbf{q}=(\sqrt{\gamma}E,\,\sqrt{\gamma}S_i)$, fluxes $\mathbf{F}$ and sources $\mathbf{S}$, which can be read off from \eqref{eq:eq_E}--\eqref{eq:eq_Si} and \eqref{eq:E_fluid_3p1}--\eqref{eq:Sij_fluid_3p1}. Given some initial data $\mathbf{p}_0|_{t=0}$ and $\mathbf{p}_1|_{t=0}$, we can construct $(\mathbf{q},\,\mathbf{F},\,\mathbf{S})$ at the initial time from \eqref{eq:E_fluid}--\eqref{eq:Sij_fluid} and evolve $(\mathbf{p}_0\,,\mathbf{q})$ until the next timestep. Then we can reconstruct $\mathbf{p}_1$ by solving \eqref{eq:E_fluid} and \eqref{eq:Si_fluid}, which is a linear system.\footnote{In fact, in 3+1 dimensions, it amounts to inverting a $4\times 4$ matrix, which can be done analytically.} Once we have $\mathbf{p}_1$ at the next time step, we can compute the fluxes and the sources (since we already have $\mathbf{p}_0$) and continue the evolution. So in this formulation the primitive recovery can be performed analytically. 

\subsection{Non-linear evolution in spherical symmetry}

In this section, we describe the formalism employed to perform fully non-linear simulations of relativistic viscous hydrodynamics with BDNK formulation. For simplicity and as proof of principle, we focus on the case of spherically symmetric spacetimes, adopting the line element  

\begin{align}
    ds^2 =&-\alpha(t,r)^2 dt^2 + g_{rr}(t,r)dr^2\nonumber \\ & +r^2g_{\theta \theta}(t,r) \left(d\theta^2+\sin^2\theta\, d\varphi^2\right) ~.
\label{isometric}
\end{align}
By defining the following first-order reduction variables 
\begin{align}
A_{r}&=\tfrac{1}{\alpha}\partial_{r}\alpha,\,{D_{rr}}^{r}=\tfrac{1}{2}g^{rr}\partial_{r}g_{rr},\,{D_{r\theta}}^{\theta}=\tfrac{1}{2}g^{\theta\theta}\partial_{r}g_{\theta\theta}\,,\nonumber\\
\partial_{t}g_{rr} &= -2\alpha g_{rr}K^{r}_{\phantom{r}r}~,~~
\partial_{t}g_{\theta\theta} =-2\alpha g_{\theta\theta}K^{\theta}_{\phantom{\theta}\theta}~,
\end{align}
and introducing the variable $\tilde{\gamma}=\sqrt{g_{rr}}\,g_{\theta\theta}$, the evolution equations \eqref{eq:eq_E}, \eqref{eq:eq_Si}, \eqref{eq:evolution_epsilon}, \eqref{eq:evolution_v} can be written as
\begin{widetext}
    \begin{align}
    \partial_t(\tilde{\gamma}\,E)+\partial_r(\alpha\,\tilde{\gamma}\,S^r )&=\alpha \,\tilde{\gamma}\big[S^r_{\phantom{r}r}K^r_{\phantom{r}r} +2S^\theta_{\phantom{\theta}\theta} K^\theta_{\phantom{\theta}\theta} - S^r\left(\tfrac{2}{r}+A_r\right)\big]\,,\\
    \partial_t(\tilde{\gamma}\,S_r)+\partial_r(\alpha\,\tilde{\gamma}\,S^r_{\phantom{r}r})&=\alpha \,\tilde{\gamma}\left[S^r_{\phantom{r}r}(D_{rr}^r-\tfrac{2}{r})+2S^\theta_{\phantom{\theta}\theta} \big(\tfrac{1}{r}+D_{r \theta}^\theta \big)-E\,A_r\right]\,, \\
    \partial_t\epsilon &=-\alpha\,\hat\epsilon\,, \label{eq:evol_epsilon_sph}\\
    \partial_t v^r &=\alpha(-\hat{\bar v}^r+K^r_{\phantom{r}r}\,v^r)\,, \label{eq:evol_vel_r_tmp}
\end{align}
\end{widetext}

As is evident from \eqref{eq:E_fluid_3p1}--\eqref{eq:Sij_fluid_3p1}, the BDNK evolution equations involve spatial derivatives of the primitive variables, i.e.,  $\partial_rv^r$ and $\partial_r \epsilon$ in spherical symmetry.  Due to shock formation during the evolution, computing these derivatives using standard finite difference methods would lead to spurious oscillations or even numerical instabilities.\footnote{The authors of \cite{Pandya:2022pif} overcame this problem by computing the derivatives using the central-WENO (CWENO) method \cite{refId0}.} To avoid this, we promote these derivatives to dynamical fields and evolve them throughout the simulation using our finite-volume numerical scheme (see Section~\ref{NM}), ensuring stable simulations.
Their corresponding evolution equations can be straightforwardly derived from \eqref{eq:evol_epsilon_sph} and \eqref{eq:evol_vel_r_tmp}:
\begin{align}
    \partial_t(\partial_r\epsilon) &=-\partial_r(\alpha\,\hat\epsilon)\,,\\
    \partial_t (\partial_rv^r) &=\partial_r\big[\alpha(-\hat{\bar v}^r+K^r_{\phantom{r}r}\,v^r)\big]\,.
\end{align}
In order to mitigate numerical instabilities during the evolution near the coordinate singularity at $r=0$, we introduce the regularised radial component of the velocity $\tilde{v}^r=\frac{1}{r}v^r$. Hence, the new evolution equations for $\tilde{v}^r$ and $\partial_r \tilde{v}^r$ are given by
\begin{align}
    \partial_t \tilde{v}^r &=\alpha\big(-\tfrac{\hat{\bar v}^r}{r}+K^r_{\phantom{r}r}\,\tilde{v}^r\big)\,, \label{eq:evol_vel_r}\\
    \partial_t (\partial_r \tilde{v}^r) &=\partial_r\big[\alpha\big(-\tfrac{\hat{\bar v}^r}{r}+K^r_{\phantom{r}r}\,\tilde{v}^r\big)\big]\,.
\end{align} 
Finally, notice that all the evolution equations are written as a system of balance laws, eq. \eqref{eq:balance_law},
where $\{\tilde{\gamma}\,E,\,\tilde{\gamma}\,S_{r},\,\epsilon,\,\partial_{r}\epsilon,\,\tilde{v}^r, \,\partial_{r}\tilde{v}^r\}$ is our final set of evolved variables in the spherically symmetric case. The primitive variables are given by $\mathbf{p}_0=(\epsilon,\,\tilde{v}^r)$ and $\mathbf{p}_1=(\hat{\epsilon},\,\hat{\bar v}^r)$ respectively, while the conserved ones are $\mathbf{q}=(\tilde{\gamma}\,E,\tilde{\gamma}\,S_{r})$. The explicit procedure to recover the primitive fields $\mathbf{p}_1 = \mathbf{p}_1 \left(  \mathbf{p}_0,  \mathbf{q}\right)$ is described in detail in Appendix \ref{con2prim_appendix}.

\subsubsection{Equation of state}

The thermodynamical quantities that describe our viscous fluid are the total energy density $\epsilon$ and the pressure $p$. The former can be expressed in terms of the rest mass density $\rho_0$ and specific internal energy $\epsilon_0$ of the fluid, namely
\begin{align}
    \epsilon = \rho_0 (1+\epsilon_0)~.
    \label{epsilon_definition}
\end{align}

From a mathematical point of view, an EoS of the form $p=p(\epsilon)$ is required to close the system of evolution equations. Physically, the EoS connects the thermodynamic variables to the microphysics that governs the fluid. Therefore, different types of matter are described by the same hydrodynamical equations, but with a different EoS.

Two commonly employed equations of state, well-suited for modelling neutron stars, can be formulated in terms of the thermodynamic variables $(\rho_0,\epsilon_0)$. The first one, mostly valid for cold neutron stars, is the polytropic EoS
\begin{align}
p=\kappa \,\rho_0^\Gamma~,
    \label{EOS:polytropic}
\end{align}
where $\kappa$ is the polytropic constant and $\Gamma$ the adiabatic index.  The second choice is the well-known ideal gas EoS
\begin{align}
    p=(\Gamma-1)\epsilon_0\, \rho_0~,
    \label{EOS:ideal_gas}
\end{align}
which is valid in more general scenarios, as it allows for the conversion of kinetic energy into thermal energy at shocks. Typical values resulting in masses and radii comparable to those of neutron stars, that we have employed in our simulations, are $\kappa=100$ and $\Gamma=2.$ 
Since our aim in this work is to study neutron stars close to their equilibrium configuration, we can construct an equation of state $p=p(\epsilon)$ by using both the polytropic and ideal gas EoS:
solving for  $\rho_0$ from \eqref{EOS:polytropic},  $\epsilon_0 \rho_0 $ from \eqref{EOS:ideal_gas}, and substituting into \eqref{epsilon_definition}, we obtain
\begin{align}
    \epsilon = \Big(\frac{p}{\kappa}\Big)^{\frac{1}{\Gamma}} + \frac{p}{\Gamma-1}~.
\end{align}
Finding solutions to this equation is non-trivial for arbitrary values of $\Gamma$.  However, for our specific choice of $\Gamma=2$, the above equation simply reduces to a quadratic equation and analytical solutions $p_{\pm}$ can be found easily.  The positive root is discarded by enforcing the physical condition $p(\epsilon=0)=0$. Therefore, our final EoS is given by
\begin{align}\label{EOS}
    p(\epsilon)=\frac{1+2\,\epsilon \,\kappa\, -\sqrt{1+4\,\epsilon\, \kappa}}{2\, \kappa}~.
\end{align}
Although this EoS might have a limited astrophysical relevance, it serves as a useful first step for implementing and testing the BDNK framework in a realistic setup.

\subsubsection{Initial data }\label{ID}

The initial data for our evolution equations requires both of geometric quantities (i.e., the spatial metric $\gamma_{ij}$, the extrinsic curvature $K_{ij}$ and the lapse function $\alpha$) and fluid variables (i.e., the energy density $\epsilon$, the fluid velocity $v^i$, together with their corresponding time derivatives $\hat{\epsilon}$, $\hat{\bar v}^i$). 
For spherically symmetric solutions modelling neutron stars, we first adopt a static metric given by the following line element in polar-areal coordinates (also known as Schwarzschild coordinates), which reads 
\begin{align}
    ds^2=-\alpha^2(R)\,dt^2+a^2(R)\,dr^2+R^2\,d \Omega^2,
    \label{schwarzschild_metric}
\end{align}
where $d\Omega^{2}=d\theta^{2}+\sin^{2}\theta\, d\varphi^{2}$ is the metric of a unit two-sphere. For the matter sector, we use the conventional perfect fluid (PF) stress-energy tensor, namely
\begin{align}
    T_{\mu\nu}=\big[p(R)+\epsilon (R)\big]u_\mu(R) \,u_\nu(R) +p(R)\,g_{\mu\nu}~.
\end{align}
Note that this choice is physically well-motivated, 
since the dissipation effects coming from the BDNK corrections vanish in the equilibrium state.
Hence, a star in hydrodynamic equilibrium should be accurately modeled by the PF stress-energy tensor. 
The next step consists of imposing the PF hydrostatic equilibrium condition $v^i=0.$  Note that as we evolve the initial data with the BDNK equations, we need to supplement the hydrostatic equilibrium condition with additional initial conditions for the time derivatives of the fluid variables;  we take them to be  $\hat{\bar v}^r=\hat{\epsilon}=0.$ Under these conditions, the Einstein and hydrodynamic equations lead to the following system of ordinary differential equations~\cite{CCC}
\begin{align}
    \frac{da}{dR} &= \frac{1+a^2(-1+8\pi  R^2 \epsilon)}{2R} a\,,\\
    \frac{d \alpha}{dR} &= \frac{-1+a^2(1+8\pi  R^2 p)}{2R} \alpha\,,\\
    \frac{d p}{dR} &= -\frac{(p+\epsilon)}{\alpha}\frac{d \alpha}{dR}
\end{align}
The above system, after providing the EoS \eqref{EOS}, can be solved numerically by imposing boundary conditions that guarantee regularity at the origin and asymptotic flatness, namely
\begin{align}
    \alpha(0) &= 1\,, \quad a(0) = 1\,,\quad p(0) = \kappa \rho_0 (0)^\Gamma\,,\\
    \lim_{R \rightarrow \infty} \alpha(R) &= \lim_{R \rightarrow \infty} \frac{1}{a(R)}\,, \ \ \ \lim_{R \rightarrow \infty} p(R) = 0\,.
    \label{boundary_conditions2}
\end{align}

Finally, once the solution for the equilibrium configuration has been found, a coordinate transformation is performed from areal-polar coordinates to maximal isotropic ones~\cite{2004PhDT.......230L}, in which the line element can be written as
\begin{align}
    ds^2 = -\alpha^2(r)dt^2 + \psi^4(r)(dr^2 + r^2 d\Omega^2)
    \label{maximal_isotropic}
\end{align}
where $\psi$ is the conformal factor.  This line element is subsequently translated into the form of the metric ansatz employed in the evolution equations, as given in Eq.~\eqref{isometric}.

\subsection{Frame Choice, Well-Posedness, and Linear Stability}\label{frame}

In this section, we introduce a set of parameters associated with the choice of frame and viscosities, and discuss the conditions they must satisfy to ensure strong hyperbolicity, causality, and linear stability of the viscous hydrodynamics equations considered here. These conditions take the form of inequalities that our parameters have to satisfy; see \cite{Bemfica:2019knx}  for a general derivation in a broader context.

Roughly speaking, a set of partial differential equations (PDEs) is strongly hyperbolic  (a subset of the well-posed ones ) if the norm of small, high-frequency perturbations remains bounded (over finite time intervals) solely in terms of the norm of their initial data. Violation of this condition would allow arbitrarily small perturbations of high frequency to grow without control, leading to the blow up of the solutions (in finite
time) from their unperturbed counterparts. This behaviour is incompatible with any physically meaningful theory, and this is precisely why establishing strong hyperbolicity, and thus well-posedness, is essential (see \cite{KreLor89, GusKreOli95, Sarbach:2012pr, Hil13} for introduction to the topic). In addition, in relativistic theories, causality requires that all physical characteristic speeds must be smaller than the speed of light, i.e., less than $1$ in natural units.

Motivated by \cite{Pandya:2022sff}, we define
\begin{equation}
    \rho\equiv\epsilon+p\,,\quad V\equiv\frac{4}{3}\eta+\zeta\,,
\end{equation}
and introduce the following parametrization 
\begin{align}
\begin{array}
[c]{ccccccc}
\eta\equiv\hat{q} 
Lc_{s}^{2}\rho\hat{\eta}
&  &
\zeta\equiv\hat{q}  
Lc_{s}^{2}\rho\hat{\zeta}
&  &
\hat{V}\equiv\frac{4}{3}\hat{\eta}+\hat{\zeta} &  & \beta_{\epsilon}\equiv
c_{s}^{2}\hat{a} 
\hat{V}L
\\
\tau_{p}\equiv\hat{s}c_{s}^{2}
L\hat{V}
&  &
\tau_{Q}\equiv\hat{a}
L\hat{V}
&  & \tau_{\epsilon
}\equiv
\hat{V}L
&  &
\label{hatted_parameters}
\end{array}
\end{align}
where $\hat{\zeta}\geq0$ and 
\begin{align}
\hat{a},\hat{q},\hat{s},L,c_{s},
\hat{\eta},\hat{V}>0. \label{hatted_parameters_positive}
\end{align} 
Apart from the speed of sound $c_s$, which is fixed by the EoS ($c_s^2:=dp/d\epsilon$), and the system's characteristic length scale $L$, these quantities are functions of the thermodynamic variables and they define the choice of frame as well as control the viscous sector of the theory.  In practice these functions are freely specified, i.e. the dimensionless hatted quantities can be functions of thermodynamic quantities;\footnote{In principle, one should be able to determine the physical transport coefficients from a calculation in the microscopic theory.  Nevertheless, from the phenomenological perspective adopted in this paper, we will treat them as free functions that can be fitted to observations.} here, for simplicity, we choose them to be constant.  The dependence on thermodynamic variables will be entirely absorbed into the functions (which may depend on $\epsilon$) multiplying the hatted quantities as we construct the variables which enter directly in the theory.
Notice that although $L$ is technically an independent parameter, for practical purposes we set $L=1$. 
On the other hand, notice that the conditions \eqref{hatted_parameters_positive} allow $\zeta$ to be zero, but $\eta$ must be strictly positive, since the latter case results in a weakly hyperbolic system, as we explain below. 

If we consider the evolution equations for our visco-fluids in a flat spacetime, the characteristic velocities (i.e., the velocities of the high-frequency modes) of the
system are $\pm c_{0},\pm
c_{\pm}$, where \footnote{Expressions \eqref{characteristic_velocities_1a}-\eqref{characteristic_velocities_1b} are obtained from the characteristic velocities written in \cite{Bemfica:2019knx} by setting the parameters to $\chi_1=\chi_2=\rho \tau_{\epsilon}$, $\chi_3=c_s^2 \rho \tau_{\epsilon}$, $\chi_4=c_s^2 \rho \tau_{\epsilon}-\zeta$, $\lambda=\rho\tau_Q$ and using \eqref{hatted_parameters}.  }%
\begin{widetext}
\begin{align}
c_{0} &  =c_{s}\sqrt{\frac{\hat{q}\hat{\eta}}{\hat{a}\hat{V}}},\label{characteristic_velocities_1a}\\
c_{\pm} &  =c_{s}\sqrt{\frac{\hat{a}\left(  1+\hat{s}\right)  +\hat{q}\pm
\sqrt{\hat{q}^{2}+\hat{a}^{2}\left(  4\hat{q}+\left(  \hat{s}-1\right)
^{2}\right)  +2\hat{a}\hat{q}\left(  1+\hat{s}\right)  }}{2\hat{a}}}.
\label{characteristic_velocities_1b}
\end{align}
\end{widetext}
To ensure that the initial value problem is well-posed, it is necessary and sufficient that $c_0$ and $c_\pm$ are real and strictly positive,  and  $c_+ \neq c_-$.

It is straightforward to verify that $c_0$ and $c_+$ are always real and positive if  \eqref{hatted_parameters_positive} is satisfied. In contrast, $c_-$ is real only if $\hat{q}$ satisfies the condition  
\begin{align}
	0 < \hat{q} < \hat{s}\,. \label{cond_1}
\end{align}
Imposing this condition and given that $\hat{a}, \hat{q}, \hat{s} > 0$, it also follows that $c_+ \neq c_-$, ensuring strong hyperbolicity.

The conditions for relativistic causality require that $c_1, c_\pm < 1$. In terms of our parametrisation, this leads to the following inequalities
\begin{align}
\hat{q} &  <\frac{\left(  1-c_{s}^{2}\right)  }{c_{s}^{2}}\frac{\left(
1-\hat{s}c_{s}^{2}\right)  }{\left(  c_{s}^{2}+\hat{a}^{-1}\right)
},\label{cond_3} \\ & \hat{s}<\frac{1}{c_{s}^{2}}. \label{cond_4} 
\end{align}
Since the characteristic speeds are written simply in term of (positive-definite) hatted parameters, these inequalities follow from a straightforward derivation.\footnote{These expressions match those in \cite{Bea:2025eov} for field redefinitions proportional to the equations of motion in a non-conformal theory, by setting $\hat{s}=1$ and identifying $\hat{q}=4/(3 a_1 c_s^2)$ and $\hat{a}=a_2/a_1$.
}

Linear stability refers to the property that small perturbations around homogeneous equilibrium solutions do not grow exponentially, but instead decay back to equilibrium over time. To derive conditions that guarantee this behavior, one linearizes the equations around equilibrium fluid configurations on a  flat spacetime, and considers plane-wave perturbations. This analysis produces inequalities that prevent such modes from growing and ensure their decay toward equilibrium, see equations (10a) and (10b) of \cite{Bemfica:2019knx} for the general conditions that ensure linear stability. In our parametrization, these conditions reduce to condition~(\ref{cond_1}), which is precisely the same requirement that guarantees $c_-$ is real. This equivalence has been verified analytically using \textit{Mathematica} \cite{WolframResearch.Mathematica1}.

On the other hand, when our viscous fluid theory is considered on an arbitrary Lorentzian background metric $g_{\mu \nu} $ (see eq. \eqref{metric}), the characteristic velocities of the fluid e $\tilde{c}_{0_{\pm}}, \tilde{c}_{1_{\pm}}, \tilde{c}_{2_{\pm}}$, are given by
\begin{widetext}
\begin{align}
 \tilde{c}_{i_{\pm}} = -\beta \cdot k + \alpha \frac{m_i W^2 (v \cdot k) \pm \sqrt{k^2 + k^2 m_i W^2 - m_i W^2 (v \cdot k)^2}}{m_i W^2 + 1}, \label{vel_w_m}   
\end{align}
\end{widetext}
where $i = 0, 1, 2$,  $m_0 = \left( \frac{1}{c_0^2} - 1 \right)$ and $m_{1,2} = \left( \frac{1}{c_{\pm}^2} - 1 \right)$. Here, $k^i$ is the wave vector (spacelike), and the quantities $v^{\mu}$ and $W$ are defined in eqs. (\ref{eq:def_u_A}) and (\ref{eq:def_W}). We will use them to select the appropriate level of ``numerical" dissipation in our setup, as we explain in the next section. 

To conclude this section, we mention that the fluid remains strongly hyperbolic under the same conditions found in the flat spacetime case; and causality and stability conditions are once again ensured by inequalities
(\ref{cond_3}) and~(\ref{cond_4}), together with the requirement that all hatted parameters remain positive. We recall that the stability condition for the fluid sector is derived via linearization around flat spacetime; it therefore remains unchanged and coincides with the previously established condition~(\ref{cond_1}).

All the conditions discussed above have been verified in our numerical implementation, as we demonstrate in the following sections.

\subsection{Numerical methods}\label{NM}
The numerical discretization is performed using the Method of Lines. The evolution equations are integrated in time through a third-order accurate Strong Stability Preserving Runge-Kutta scheme, with a Courant factor $\Delta t/\Delta r = 0.25$ such that the Courant-Friedrichs-Levy (CFL) stability condition is satisfied. 
The spatial discretisation is based on a third-order finite-volume scheme, which is equivalent to a fourth-order finite difference scheme with third-order dissipation (this algorithm is also known as the ``finite-difference Osher-Chakrabarthy'' scheme, or FDOC in short)~\cite{Alic:2007ev,Bona:2008xs,Palenzuela:2018sly,CCC}.  Since we are interested in understanding the effect of physical viscosity on NSs, we take special care in choosing the maximum characteristic speed from the six characteristic velocities of the viscous BDNK fluid (see Eq.~\eqref{vel_w_m}), which determines the amount of ``dissipation'' used by the FDOC scheme. We also set a minimum velocity $0.1 c$ to ensure that a small amount of numerical dissipation is present near the surface of the star and in the atmosphere, which helps to stabilize our simulations.  We identify the atmosphere as the set of points where the local pressure satisfies the condition $p<\kappa \rho_{\text{0,atms}}^\Gamma$ ($\rho_{\text{0,atms}}=10^{-12} M_{\odot}^{-2}$ is used in our simulations).  These points will be provided with an identical constant value for the baryonic density ($\rho_0 = 10 ^{-13} M_{\odot}^{-2}$), which is then used to update the relevant variables (e.g., $\epsilon$, $p$) in the atmosphere.  The velocity of the fluid and the time derivatives $\hat{\epsilon}$, $\hat{\bar v}^r$ will be set to zero there.

We perform simulations across a range of high-resolutions  $\Delta r=[0.001 - 0.0032] M_\odot$.  To avoid any potentially problematic behaviour at the origin $r=0$, we adopt a staggered grid in our simulations.  The box size is limited to $r_{\text{max}}=20 M_\odot,$ which is sufficient for our current simulations on a fixed background as we explain in the next Section~\ref{simuNS}. However, we have verified that by changing the position of the outer boundary, the results do not
vary significantly. We use outflow boundary conditions for the fluid matter fields at the outer boundary. 

\section{Numerical evolution of stable neutron stars}\label{simuNS}

In this Section we present the results of long-time numerical simulations of NSs using the evolution formalism of BDNK and the numerical setup described in the previous section.  For simplicity, the fluid is evolved on a fixed background spacetime (i.e., the commonly known Cowling approximation \cite{10.1093/mnras/101.8.367}), which is set to be the curved geometry of the NS initial data.  We note that while the Cowling approximation is a standard simplification used in NS studies (see e.g. \cite{Thierfelder:2011yi,Font:2001ew, Yoshida_1999, Counsell:2024pua}), the predictions obtained under this assumption differ from evolving the fully coupled system due to its suppression of certain dynamical channels (such as gravitational backreaction of the fluid).

The existence of multiple parameters in the theory that govern the hydrodynamic frame and the physical viscous contributions complicates a systematic study of these effects.  To address this issue, we first introduce a concrete framework to explore the parameter space built upon the hyperbolicity requirements studied in Section~\ref{frame}. By using this approach, we select four representative sets of parameters to perform simulations of stable NS configurations. 

The initial data for the NS is constructed following the procedure presented in Sec~\ref{ID}. In particular, we will construct an equilibrium configuration with central rest mass density $\rho_{0,c}=0.00128 M_{\odot}^{-2}$ (or equivalently $\epsilon_c=0.00144 M_{\odot}^{-2} $) and total gravitational mass $M_{\rm{T}}=1.4 M_{\odot},$ which has been extensively studied both analytically and numerically~\cite{CCC,Font:2001ew}. We show that long-time stable evolution of such a star, employing viscous hydrodynamics, is possible for our four representative cases. Finally, we study numerically the quasi-normal mode (QNM) spectrum of the NS together with the decay rate of the fundamental mode, analysing its dependence with respect to the frame and viscosity parameters. 

\subsection{Choice of theory parameters}
The BDNK formulation is fully determined by specifying the free functions $\tau_{\epsilon},\tau_{p},\tau_{Q},\eta,\zeta$, 
with the first three parameters fixing the frame and the latter two fixing the viscosity.  In Section~\ref{frame}, we have re-expressed these functions in terms of $(\hat{a}, \hat{q}, \hat{s},
\hat{\eta}, \hat{\zeta})$.  We will now outline our method in choosing the values of the hatted quantities for our numerical simulations.
In order to ensure that well-posedness and causality conditions are satisfied, we first consider fixing the parameters $(\hat{s} ,\hat{a},\hat{q})$.
The choice $\hat{s}=1$ is used to simplify the frame parametrization, leading to $\tau_p=c_s^2 \tau_{\epsilon}$.  
We set $\hat{q}=0.999$, very close to the upper bound, in order to maximize the allowed range physical viscosity. Finally, we set $\hat{a}=1$ for simplicity, obtaining $\tau_{Q}=\tau_{\epsilon}$. Notice that with these choices, fixing $\tau_{\epsilon}$ completely determine the frame. In addition, these values of $\hat{q}$ and $\hat{a}$ satisfy the well-posedness condition \eqref{cond_1} and the relativistic causality condition \eqref{cond_3}, as long as $c_s^2  < 1/3$, a bound that is satisfied in our simulations. For these parameters, the characteristic velocities are given by
\begin{align*}
c_0 &= 0.9995 \sqrt{\hat{\eta}/\hat{V}} c_s,\quad c_{+}= 1.732c_s,\nonumber\\
& \qquad c_{-} = 0.0183 c_s~. 
\end{align*}
Once $(\hat{s},\hat{a},\hat{q})$ are fixed, we proceed to fix the frame by specifying $\tau_\epsilon$; finally  we  choose the remaining parameters $(\hat{\eta}, \hat{\zeta})$ such that they are consistent with our choice of $\tau_\epsilon$.  Empirically, we find that stable evolutions can be achieved imposing the condition 
\begin{align}
\tau_{\epsilon}=\tfrac{4}{3}\,\hat{\eta}+\hat{\zeta} \lesssim 0.1~.  
\end{align}
This condition was found by performing high-resolution simulations (with $\Delta r=0.001 M_\odot$) of NSs over a range of values of $\tau_{\epsilon}$\footnote{We have evolved larger values of $\tau_\epsilon$ using a lower resolution.  However, we will not consider these cases here as they are not stable in high-resolution simulations.}. We use values within this range. 
Indeed, if the BDNK formulation is to be interpreted as an EFT, then the relaxation time $\tau_\epsilon$ must be sufficiently small to ensure that the system remains within the regime of validity of the theory.

Note that this is just a guiding principle for us to systematically search for a set of suitable parameters that allow us to perform stable evolutions. In practice, we have found certain values of the parameters that satisfy all the above conditions and yet they still lead to unstable  numerical evolutions.
We will avoid such cases in our following analysis, and focus on the parameters where stable evolution is possible with our highest resolution. 

We consider four cases:
\begin{itemize}
    \item \texttt{smallSB-F2}: A case with small viscosity both in the shear and bulk viscosity in a given frame, corresponding to the parameter choice $(\tau_\epsilon, \hat{\eta} 
    , \hat{\zeta} 
    ) = (0.023,0.01,0.01)$.
    \item \texttt{medS-F2}: A case with medium viscosity contribution coming only from the shear viscosity, obtained with the parameters $(\tau_\epsilon, \hat{\eta}
    , \hat{\zeta}
    ) = (0.023, 0.01725, 0)$. It employs the same hydrodynamic frame as \texttt{smallSB-F2}.
    \item \texttt{highB-F9}: A case with high viscosity, predominantly originated from the bulk viscosity, obtained with the parameters $(\tau_\epsilon, \hat{\eta} 
    , \hat{\zeta} 
    ) = (0.092, 0.0015,0.09)$.
    \item  \texttt{medSB-F9}: A case with half the bulk viscosity of \texttt{highB-F9} while keeping the same hydrodynamic frame, such that the shear and bulk are comparable. It corresponds to the parameter choice $(\tau_\epsilon, \hat{\eta} 
    , \hat{\zeta} 
    ) = (0.092, 0.03525,0.045)$.
\end{itemize}

We emphasise that we chose viscosities ($\eta,\zeta$) with energy dependence given by $c_s(\epsilon)^2(\epsilon+p(\epsilon))$, see \eqref{hatted_parameters}, which can be roughly thought as growing near linearly in $\epsilon$ in the relevant regime of our simulations. 
To explicitly show the units, let us consider for example the case smallSB-F2: $\eta=0.00999 c_{s}^{2}(\epsilon+p)$ $M_{\odot}$, with $\epsilon,p$ in $M_{\odot}^{-2}$. Or in SI units: $\eta=9.12\times10^{21} c_{s}^{2}(\epsilon+p)$ $Pa \cdot s$, where $\epsilon,p$ are in $J/m^3$ and $c_s$ in $m/s$. 

Finally, we would like to note that the physical transport coefficients may, in principle, be determined using a microscopic theory, and efforts have been made within the community towards computing such coefficients (e.g. \cite{Alford:2020lla,Alford:2022ufz,Hernandez:2024rxi}).  However, we have not used them as our guide in choosing our parameter values because of the different assumptions in our work.  Moreover, we adopt a more phenomenological approach in this work by treating the transport coefficients as free functions that can be fitted to observations\footnote{
This approach is standard in quark-gluon plasma physics, where the transport coefficients 
$\eta$ and $\zeta$ are extracted from fits to experimental data
\cite{Heinz:2004qz,Romatschke:2017ejr,Parkkila:2021tqq}.}.  The fitted values will then help us constrain the microscopic theory.  This motivates the numerical search of theory parameters outlined in this Section.

\subsection{Qualitative features of the simulations}
\begin{figure}
    \centering
    \includegraphics[width=1\columnwidth]{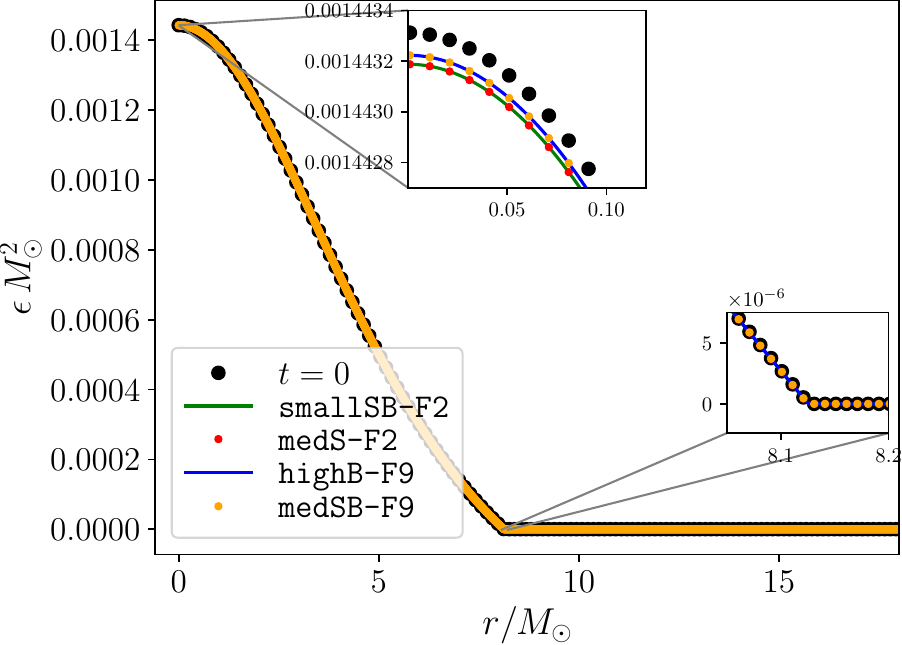}
    \caption{\emph{Comparison of late time configurations of $\epsilon (r)$ for different viscous cases.}  The radial profile of the energy density $\epsilon (r)$, comparing the initial data and the late-time evolution profiles at $t=8000 M_\odot$ for the four cases we considered with resolution $\Delta r=0.002 M_\odot$.  The two insets emphasise the behaviour near the centre and the surface of the star, showing slight deviations due to numerical dissipation and truncation errors. The use of a different hydrodynamic frame might also result in small differences in the late-time configurations.}
    \label{fig:stable_tau_comparison}
\end{figure}

Using the choices of parameters above, we perform the evolution of the stable neutron star
until $t_{f}=8000 M_\odot$ (except for the highest resolution case $\Delta r=0.001 M_\odot$ which reached $t_{f}=4500 M_\odot$ due to its  computational cost). Initially, the star is not perturbed, apart from numerical discretisation errors. By monitoring the evolution of the fluid fields, we confirm that stable evolutions can be achieved over a wide range of viscosities within the BDNK formulation. For instance, the initial and late-time (i.e., at $t_{f}=8000 M_\odot$) profile of the energy density $\epsilon$ is displayed in Fig.~\ref{fig:stable_tau_comparison}. At late times, a careful examination (i.e., see the inset of Figure \ref{fig:stable_tau_comparison}) indicates that $\epsilon(r)$ near the centre and the surface of the star exhibits slight deviations due to numerical dissipation and truncation errors for all the cases, which should vanish in the continuum limit.  The fact that we are using different hydrodynamic frames might also lead to small deviations in the late-time configuration.

Along these lines, we may also explicitly verify that the bound $c_s^2 <1/3$ is satisfied in our simulations.  By computing $c_s^2=dp/d\epsilon$ and substituting the values of $\epsilon$ reached within the NS into the expression, we find that the maximum speed occurs at the center of the star with $c_s \approx 0.45$, which decreases monotonically as we approach the surface of the star.  Given the stability of the NS across all cases, this bound remains satisfied throughout our simulations.

\begin{figure}
    \centering
    \includegraphics[width=1\columnwidth]{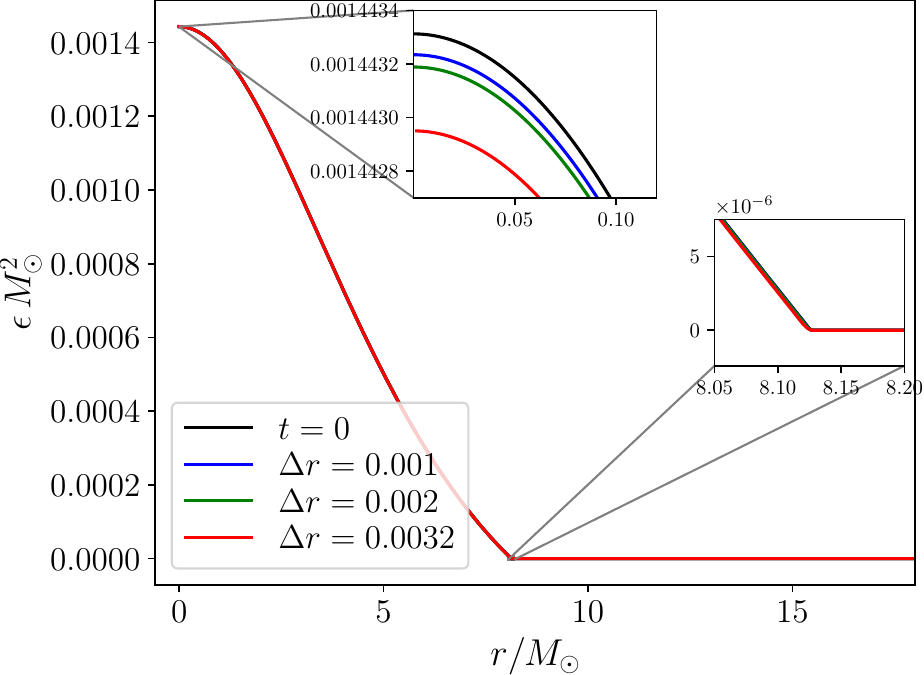}
    \caption{\emph{Comparison of late time configurations of $\epsilon (r)$ across different resolutions.}  The radial profile of the energy density $\epsilon (r)$ across different resolutions for \texttt{smallSB-F2} at $t=4500 M_\odot$.  The insets show  the behaviour near the centre and the surface of the star, with the former inset demonstrating qualitative convergence.}
    \label{fig:stable_dr_comparison}
\end{figure}

In Fig.~\ref{fig:stable_dr_comparison} it is displayed the same profile $\epsilon(r)$ at late times across multiple resolutions for the case \texttt{smallSB-F2}. We observe a deviation from the constant stationary value due to numerical errors, which decreases as the resolution is increased, showing that the star remains stable. This reflects, at a qualitative level, the convergence of our simulations. A detailed quantitative discussion about convergence is presented in  Appendix~\ref{covergencesection}. 

We would like to note that our solutions are well in the EFT regime of hydrodynamics, or in other words, where the first order approximation is reliable. This can be demonstrated, for example, by estimating the dimensionless number $\zeta \nabla_\mu u^\mu/p$ at the centre of the NS at late times.  For the highest bulk viscosity case we considered (\texttt{highB-F9}), we find that $ \nabla_\mu u^\mu \sim \partial_rv^r \approx 10^{-5} M_\odot^{-1}$, $\zeta \approx 10^{-5}M_\odot^{-1}$ and $p\approx10^{-4}M_\odot^{-2}$.  This implies $\zeta \nabla_\mu u^\mu/p \approx 10^{-6}$, which demonstrates that we are in the regime of validity of first order viscous hydrodynamics.

\subsection{Radial oscillations}

A perturbed neutron star undergoes characteristic oscillations that can be described in terms of a set of quasi-normal modes (QNMs). These modes can be extracted by analysing the oscillations in the central energy density of the NS, which are excited by the discretisation errors inherent in our numerical scheme \cite{Font:2001ew,Baiotti:2008nf,Valdez-Alvarado:2012rct}. Our assumption of spherical symmetry only allows us to study the radial modes. In this section, we compute the QNM frequency spectrum and decay rate of the fundamental modes predicted by the BDNK formulation for  different frame choices and viscous parameters, and compare them with the PF model.

\subsubsection{QNM frequency}

\begin{figure*}
    \centering
    \includegraphics[width=0.99\linewidth]{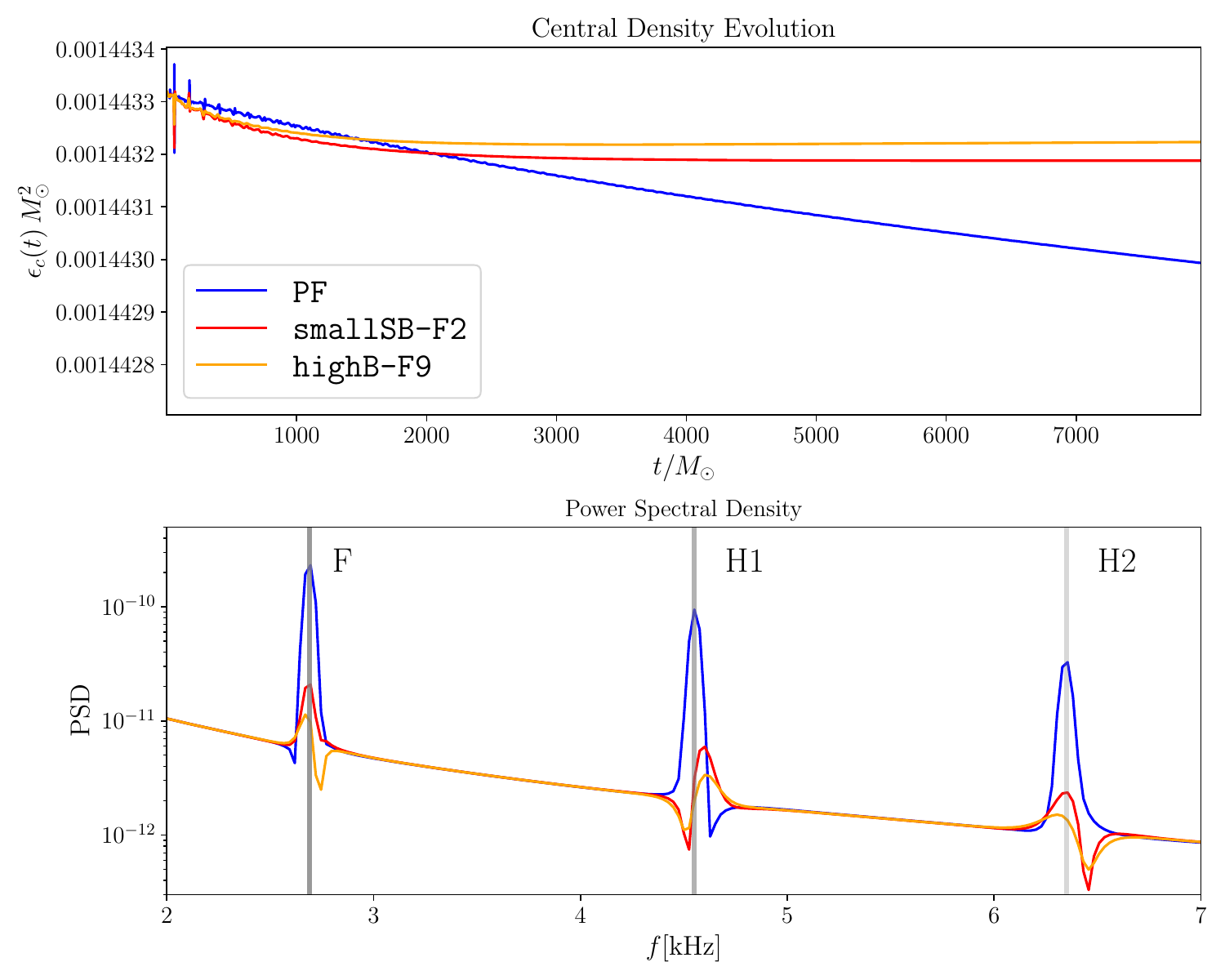}
    \caption{\emph{Central density oscillations and QNM spectrum.} \textbf{Top panel:} The oscillations in the central energy density $\epsilon_c (t)$ as a function of time, comparing the three cases: perfect fluid (\texttt{PF}), BDNK with small viscosity (\texttt{smallSB-F2}), and BDNK with large viscosity (\texttt{highB-F9}).  These simulations have a resolution of $\Delta r=0.002 M_\odot$ and the value of $\epsilon_c$ is extracted every $\Delta t=1 M_\odot$ up to $t=8000 M_\odot$. \textbf{Bottom panel:} The power spectral density (PSD) of the NS QNMs as modeled by the BDNK formulation and the perfect fluid.  The PSD is obtained by performing a Fourier transform using a Blackman window on the data shown in the top panel.  While the $f$-mode (F) appears to be consistent across all three cases, the viscous effects appear to have a small impact on modifying the overtones (H1, H2).}
    \label{fig:QNM_raw_data}
\end{figure*}

We study the power spectrum of the NS oscillations using the long-time (i.e. $t_f=8000 M_\odot$) evolutions of the PF and the BDNK cases \texttt{smallSB-F2} and \textbf{\texttt{highB-F9}}. The central value $\epsilon_c$ as a function of time is displayed for these three cases in the top panel of Fig. \ref{fig:QNM_raw_data}. The perturbation induced by the discretization errors is very small and only discernible by eye in the first $t\lesssim 1000 M_\odot$. In the bottom panel of Fig. \ref{fig:QNM_raw_data} we display the power spectral density (PSD) obtained from computing the Fourier transform of the central energy density using a Blackman window for the cases shown in the top panel. 
This plot shows that in our current setup, i.e., spherical symmetry and the Cowling approximation, the regime of viscosities explored does not significantly impact the frequencies of the QNMs.  In particular, the fundamental mode (also known as the $f$-mode) frequency is consistent across all three cases (see Table \ref{QNM_freq_table}), while the frequencies of the overtones appear to have a slight dependence on the viscosity.  A possible explanation for this observation is that viscous effects operate at short length scales since the viscous terms arise at higher orders in the gradient expansion, hence they modify the short distance physics. Therefore, viscous terms should have a greater effect on the high frequency modes since those are more sensitive to short length scale modifications.  The $f$-mode, being the lowest frequency mode, likely operates at a longer length scale and hence it should be less sensitive to the details of viscous corrections \cite{Luis_comment}. 
The value that we obtain matches the expected $f$-mode frequency reported in the literature \cite{Thierfelder:2011yi,Font:2001ew} reasonably well under the Cowling approximation. The frequencies of the fundamental mode (F) and the first two overtones (H1,H2) in each case are listed in Table \ref{QNM_freq_table}.

We note, however, that these conclusions may change in more generic scenarios.  For example, shear viscosity might play a stronger role in changing the QNM spectrum if we do not restrict ourselves to spherical symmetry. Beyond the Cowling approximation, the interaction between gravity and fluid dynamics may also produce a richer frequency spectrum.  These possibilities will be studied in more detail in a future work.

\begin{table}[!h] 
\begin{tabular}{c | c c c} 
 \hline
 Mode & \texttt{PF} & \texttt{smallSB-F2} & \texttt{highB-F9} \\ [0.5ex] 
 \hline\hline
 F & 2.69 & 2.69 & 2.67 \\ 
 \hline
 H1 & 4.55 & 4.60 & 4.60 \\
 \hline
 H2 & 6.36 & 6.36 & 6.30 \\
 \hline
\end{tabular}
\caption{\emph{{Frequency of the fundamental mode and overtones in kHz.}} While the $f$-mode appears to be largely consistent across all cases, the overtones show a slight dependence on viscosity.}
\label{QNM_freq_table}
\end{table}

\subsubsection{QNM decay rate}\label{QNM_decay_rate_section}
In addition to the frequency spectrum, in viscous fluids,  QNMs present also dissipative effects that are characterised by their decay rates.  Here, we focus on extracting the decay rate of the $f$-mode. In order to avoid contamination from other overtones, we only analyse the data at late enough times such that the higher overtones have decayed away sufficiently~\cite{Chabanov:2023abq,Radice:2011qr,Cerda-Duran:2009xon}.  We then compare the BDNK decay rates with decay rates rates obtained with the PF, which arise from the inherent numerical dissipation of our method.  For the purpose of this section, we will consider PF simulations with end time $t_f=20000 M_\odot$ as the lack of physical dissipation makes the decay rate extraction especially challenging, while the BDNK simulations terminate at $t_f=8000 M_\odot$.

\begin{figure}
    \centering
    \includegraphics[width=1\columnwidth]{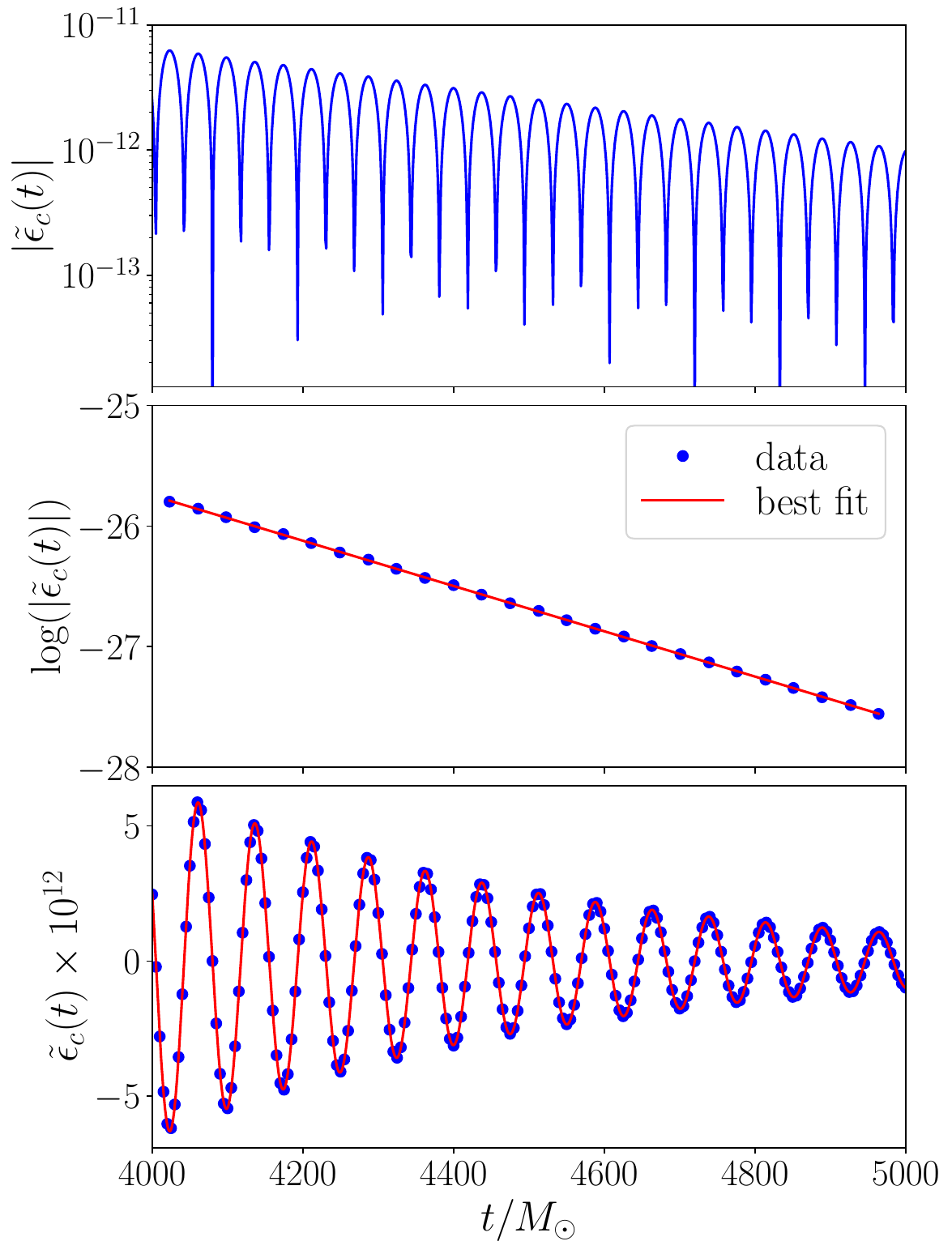}
    \caption{\emph{The different stages of the decay rate extraction procedure, demonstrated using the data from case \texttt{smallSB-F2}}.  In the top panel, we show the absolute value of the filtered data $\tilde{\epsilon}_c$ plotted against time.  Note that the signal is decaying exponentially at late times.  In the middle panel, we plotted the maximum of $\log (|\tilde{\epsilon}_c|)$ along with the best fit line, which provides an estimate for the decay rate.  In the bottom panel, $\tilde{\epsilon}_c$ is fitted with a damped sinusoidal, from which we recover the decay rate and the frequency of the $f$-mode.}
    \label{fig:QNMdecay}
\end{figure}

To extract the decay rates, we adopt the method proposed in \cite{Chabanov:2023abq}. The result of the fitting procedure is displayed in Fig. \ref{fig:QNMdecay}. First, a Butterworth filter of order four is applied to the raw data $\epsilon_c$ to clean the signal.  By filtering the low-frequency content in the data, we remove the global drift and obtain a ``clean'' signal $\tilde{\epsilon}_c$ that oscillates around zero.  Informed by the QNM frequency spectrum, we choose our frequency cutoff window to be $[0.01, f_\text{sampling}/10]$ (in code units $1/M_\odot$) so that the fundamental mode is well covered within the frequency window, and the low-frequency modes are appropriately removed.  By plotting the filtered data on a logarithmic scale, we can identify a range in time in which the signal is observed to be exponentially decaying (which manifests itself as a straight line in a log-scale plot, see the top panel of Fig. \ref{fig:QNMdecay}).  We interpret the data within this time range as being dominated by the $f$-mode.  A fitting window which covers part of this range can then be appropriately chosen to extract information about the $f$-mode.
Within this fitting window, we take the logarithm of the absolute value of the filtered data and extract the maxima, which can then be fitted with a straight line.  The slope of the fit gives the  decay rate of the $f$-mode (see middle panel in Fig. \ref{fig:QNMdecay}). 

In order to verify the results obtained with the above procedure, we also perform a non-linear least squares fit using the following function: 
\begin{equation}
    \tilde{\epsilon}_c(t) = A \exp\left(-\frac{t}{\tau}\right)\cos(\omega t+\phi_0)+C
\end{equation}
where the damped sinusoid provides a fit to the fundamental mode, and the constant is introduced to take care of any potential residual numerical noise in the data.
This fitting method allows us to check that the fitted frequency indeed agrees with the expected value for the fundamental mode as obtained from the Fourier transform, and gives us an independent way to estimate the decay rate
(see bottom panel of Fig. \ref{fig:QNMdecay}).  To verify that our fitted results are robust, we perform the fitting procedure with different fitting windows to ensure that our fitted decay rates do not depend on the window.  The starting time of the window is selected based on both the resolution and the viscosity under consideration.  In general, the smaller (higher) the viscosity (resolution), the later we start the fitting.~\footnote{For high viscosities, the oscillations decay more rapidly and the amplitude of $\tilde{\epsilon}_c$ reaches machine precision level sooner than the small viscosity cases, preventing us from fitting at late times.}

Using the above methods, we extract the decay rates for our four cases using the simulations with resolution $\Delta r=0.002 M_\odot$.  The results are summarized in Table \ref{QNM_decay_table}, with $1/{\tau}_\text{l}$ denoting the decay rate obtained from linear fitting, $1/{\tau}_\text{nl}$ from nonlinear fitting, and $\omega_{\text{nl}}$ the frequency estimated from the nonlinear fitting.  

We find $\omega_{\text{nl}}$ to be close to the value obtained from the Fourier transform. The extracted angular frequency $\omega_{\text{nl}}=0.0834 M_\odot^{-1}$ translates to $f=2.71$ kHz in physical units, suggesting that we did not over-filter the raw data when applying the Butterworth filter.  The computed decay rates do not exhibit a significant dependence on the fitting window, provided that the window does not cover the data at sufficiently early or late times, which is necessary to avoid possible contamination  arising from edge effects of the Butterworth filter and the overtones at early times.  Furthermore, the decay rates obtained from both fitting methods agree with each other to a large extent, which reflects the robustness of our results.\\

\begin{table}[!h] 
\begin{ruledtabular}
\begin{tabular}{c | c c c} 
 \hline
 Case & $\frac{1}{\tau_\text{l}}$ & $\frac{1}{\tau_\text{nl}}$ & $\omega_\text{nl}$ \\ [0.5ex] 
 \hline\hline
 \texttt{smallSB-F2} & $0.00157$ & $0.00157$ & $0.0834$ \\ 
 \hline
 \texttt{medS-F2} & $0.00150$ & $0.00150$ & $0.0834$ \\
 \hline
 \texttt{highB-F9} & $0.00215$ & $0.00215$ & $0.0834$ \\
 \hline
 \texttt{medSB-F9} & $0.00182$ & $0.00182$ & $0.0834$ \\
 \hline
\end{tabular}
\end{ruledtabular}
\caption{\emph{Decay rates of the $f$-modes in code units ($M_\odot^{-1}$).}  This table compares the decay rates obtained from linear and non-linear fitting at $\Delta r =0.002 M_\odot$, in which a close agreement is found (the errors only affect the last significant figure of the presented values by at most $\pm 0.00001$).  For all cases, the fitted frequency $\omega_{\text{nl}}$ is found to be $0.0834 M_\odot^{-1}$;  this translates to a frequency of $f=2.71$ kHz in physical units, which agrees well with the results of the Fourier transform.}
\label{QNM_decay_table}
\end{table}

Even though the above fitting accurately determines the decay rates from our simulations, there is one remaining problem to be addressed.  Due to the discretisation of the evolution equations, there will always be a source of numerical viscosity affecting our system.  The numerical viscosity decreases with increasing resolution, and should vanish in the continuum limit.  Because of this, the decay rates obtained above are  a result of the combined effect of numerical and physical viscosity present on our simulations.  To disentangle the effects of numerical viscosity from physical viscosity, we perform simulations in different resolutions and extrapolate the decay rates to the continuum limit using an equation proposed by \cite{Chabanov:2023abq}, where it was used to estimate the physical bulk viscosity:
\begin{align}\label{extrapolation_equation}
    \frac{1}{\tau_{\Delta r}} =\frac{1}{\tau_0}+m(\Delta r)^p
\end{align}
with $\frac{1}{\tau_{\Delta r}}$ being the extracted decay rate, $\frac{1}{\tau_0}$ the continuum decay rate, and $m(\Delta r)^p$ reflects the numerical contribution to the decay rate.  The value of $p$ is expected to be compatible with the convergence order of our results.

For this purpose, we consider a series of simulations with resolutions $\Delta r=[0.002,0.0024,0.0028,0.0032] M_\odot$. Then, following the  procedure outlined above, we estimate the continuum decay rate using \eqref{extrapolation_equation}.  The  value used in the extrapolation for each case/resolution is obtained by averaging the decay rates obtained from varying the fitting method/fitting window and taking two significant figures of the average.  As a result of the fitting, one obtains an estimate of the continuum decay rate $\frac{1}{\tau_0}$ and the convergence order $p$.  For completeness, we also present the extrapolation results for the PF.  Unlike the BDNK formulation, the PF does not have any dissipative effects at the continuum level.  Therefore we will set $\frac{1}{\tau_0}=0$ explicitly to avoid inaccuracies in our fitting.

\begin{table*} 
\begin{ruledtabular}
\begin{tabular}{c || c c c c c} 
 $\Delta r/M_\odot$ & \texttt{PF} & \texttt{smallSB-F2} & \texttt{medS-F2} & \textbf{\texttt{highB-F9}} & \texttt{medSB-F9}\\ 
 [0.5ex] 
 \hline\hline
 0.0032 & 0.00023 & 0.0019 & 0.0018 & 0.0024 & 0.0021\\ 
 \hline
 0.0028 & 0.00021 & 0.0018 & 0.0017 & 0.0024 & 0.0020\\
 \hline
 0.0024 & 0.00019 & 0.0017 & 0.0016 & 0.0023 & 0.0019\\
 \hline
 0.0020 & 0.00018 & 0.0016 & 0.0015 & 0.0022 & 0.0018\\
 \hline
 0 (extrapolated) & NIL & 0.0011 & 0.0010 & 0.0017 & 0.0013\\
 \hline\hline
 0 (extrapolated)  $[s^{-1}]$ & NIL & 220 & 200 & 350 & 260\\
\end{tabular}
\caption{ {\em Decay rates of the fundamental modes as a function of resolution.}  This table shows the measured decay rates $\frac{1}{\tau_{\Delta r}}$ of the $f$-mode as a function of resolution for \texttt{PF} as well as the four BDNK cases we considered, and their corresponding extrapolated values $\frac{1}{\tau_0}$.  All values are measured in code units except the ones in the final row, which represents the extrapolated decay rates in physical units $s^{-1}$.  Note the \texttt{PF} does not have an extrapolated value since we assumed $\frac{1}{\tau_0}=0$ in this case.
}
\label{tab:QNM_table} 
\end{ruledtabular}
\end{table*}

\begin{figure}
    \centering
    \includegraphics[width=1\columnwidth]{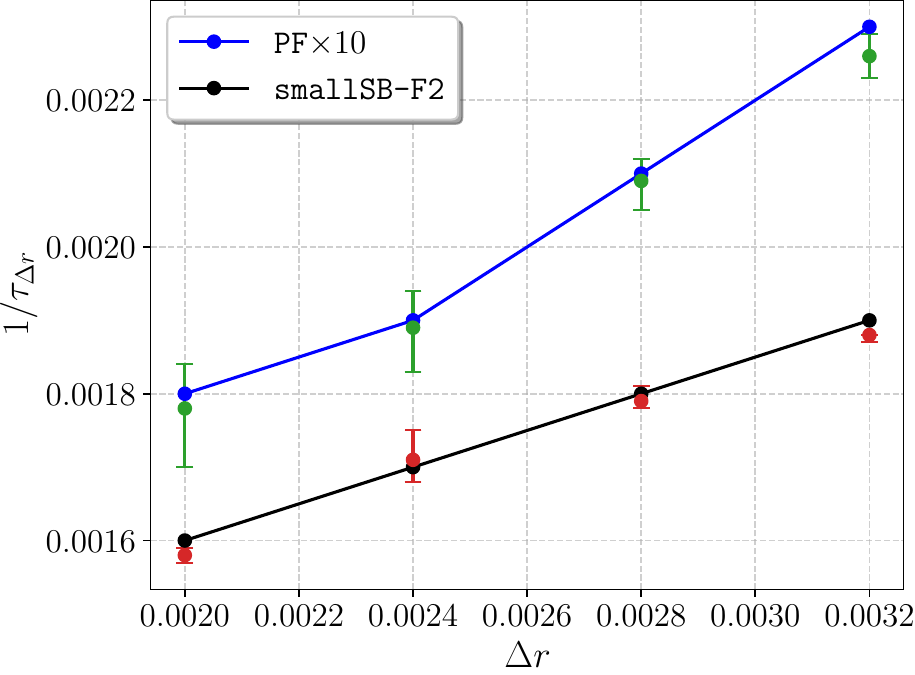}
    \caption{\emph{Decay rate (in code units) as a function of resolution}.  The cases \texttt{smallSB-F2} and \texttt{PF} (scaled by a factor of 10) are shown to demonstrate the resolution dependence.  The red and green dots are the measured values of decay rates accurate to 3 significant figures, with the error bar representing the variations one obtain by varying the fitting method/window.  To perform the extrapolation, the measurements (for a given resolution) are averaged and taken to be correct to 2 significant figures (represented by the black and blue dots).  The extrapolated curve is not shown here.}
    \label{fig:QNMextrap}
\end{figure}
The results of the extrapolation procedure are outlined in Table \ref{tab:QNM_table}.  For all BDNK cases, we find marginal convergence in $p$ (i.e., $p=1$) \footnote{Note that the extrapolation for \texttt{highB-F9} is done using only the data points from the three highest resolutions so that we can find convergence.}.  For \texttt{PF}, even though the data shows good convergence at first sight (see Fig. \ref{fig:QNMextrap}), the best fit yields $p=0.54$.  However, one should approach the quantitative decay rates obtained from \texttt{PF} with caution:  due to the lack of a physical damping mechanism, the \texttt{PF} simulations never reach a state where it is only dominated by one single mode within the simulation time we explored.  The situation worsens in the high-resolution simulations, as the numerical dissipation weakens with increasing resolution.  Therefore, our fitting procedure (which assumes the presence of one dominant mode) might not faithfully capture the decay rate of the $f$-mode as accurately as the BDNK cases.  Nevertheless, the \texttt{PF} simulations still provide us with an estimate of the strength of the numerical viscosity, which can be compared against the BDNK cases.

Our results agree with the expectation that the high viscosity cases have a larger decay rate than the small viscosity cases, and the decay rates for the viscous cases are much larger than the ones obtained from the PF model.  However, there are also unexpected elements in our analysis:  comparing the decay rates of \texttt{smallSB-F2} and \texttt{medS-F2}, we find that pure shear viscosity contribution to the decay rate is comparable to the cases with higher bulk viscosity, suggesting that it may have a non-negligible effect even in spherical symmetry.  Nevertheless, the bulk viscosity  still plays the leading role in determining the decay rate, as can be seen by comparing \texttt{highB-F9} with \texttt{medSB-F9}: increasing bulk viscosity at the expense of decreasing shear viscosity (to allow for comparisons within the same hydrodynamic frame) produces a net increase in the decay rate.

At the present stage, we focus on exploring configurations that are close to equilibrium (enforced by our choice of EoS), which limits the range of viscosities we can explore.  This is reflected in the similar values of the measured decay rates across all of our cases.  By using a more general EoS, we anticipate that a wider range of viscosities could be explored, possibly leading to more variations in the decay rate.

\section{Discussion}
The recent breakthrough in gravitational wave astronomy has provided us with a new tool to probe the universe, allowing us to take a glimpse into the strong gravity regime.  Of all the possible events that can generate gravitational waves, binary NS mergers are amongst the most fascinating ones, as they probe not only the dynamics of the strong field regime of gravity but also the properties of matter in extreme conditions.  In order to accurately extract the information encoded in the gravitational waves, one need theoretical models that include all relevant physics that could impact the dynamics of the system.  With recent studies suggesting the potential importance of viscous effects in the post-merger regime of a binary NS system, it is of utmost interest to investigate the possibility of incorporating viscous effects into our models of NS binary mergers.

This motivates us to apply relativistic viscous hydrodynamic theories to model NS binary mergers.  In particular, the recently proposed BDNK formulation of relativistic viscous hydrodynamics is particularly attractive for these applications.  It is known to be causal, stable, locally well-posed, strongly hyperbolic, and satisfies the second law of thermodynamics (in the regime of validity of the theory), giving it an edge over the currently used MIS formulation,  which is known to suffer from causality problems.  Therefore, as a first step towards achieving this goal, we performed simulations of spherically symmetric NS configurations using the BDNK formulation on a fixed spacetime background, adopting a simplified model for the EoS and transport/viscous parameters.

In order to systematically explore the parameter space of the fluid sector, we adopted a scheme that allowed us to consistently choose the magnitude of the viscous contributions for a fixed hydrodynamic frame.  Using the parameters chosen with this scheme, we found that stable evolutions can be achieved up to late times within a subset of possible hydrodynamic frames.  By considering the numerical perturbations in the same simulations, we were able to extract the frequency spectrum of the QNMs of the oscillations in the central energy density of the NS.  Comparing the frequency of the leading QNM obtained with a PF and BDNK, with high viscosity and low viscosity, we found no significant effects of the viscosity.  It is possible that this is a consequence of some of the assumptions we made in this study (e.g., spherical symmetry, Cowling approximation, or the EoS), and this conclusion should be revisited in a future work once some of the assumptions are relaxed.  Furthermore, by considering the data at late times where the decay in perturbations is dominated by the $f$-mode, we extracted the decay rate of the $f$-mode from the signal.  In order to disentangle the effects of physical viscosity and numerical viscosity, we performed the decay rate extraction procedure for the same physical setup across different resolutions, and extrapolated the decay rates as a function of resolution towards the continuum limit to estimate the decay rate caused by the physical viscosity alone.

There are different avenues through which this work can be extended in the future. First, in order to consistently model neutron stars described by the BDNK fluid coupled to GR in full generality, one must drop the assumption of spherical symmetry and relax the Cowling approximation and evolve the spacetime geometry dynamically.   As another step towards more realistic simulations, it is also necessary to evolve the baryon density independently by enforcing baryon number conservation.  This allows for the possibility of using a more general EoS, which is necessary to capture the realistic out-of-equilibrium effects in NSs.  Under these generalizations, one may even try to explore a wider range of viscosities than the ones considered in this work.  Finally, one can also include additional relevant physics into the model that has been neglected so far in our study, such as magnetohydrodynamics (MHD) and microphysics.  These improvements will pave the way for more realistic models of binary NSs mergers, which may be used to fit observational data by varying $\eta, \zeta$ for a given EoS in order to constrain the transport coefficients, and eventually get a better understanding of the properties of extreme matter.

\section{ACKNOWLEDGMENTS}

We thank Luis Lehner and Marcelo Rubio for their enlightening discussion about hydrodynamics.  We also thank Óscar H. Petit for his comments in the first version of the manuscript.
HLHS is supported by the Croucher Scholarship.
MB acknowledge partial support from the STFC Consolidated Grant no. ST/Z000424/1.
YB acknowledges support by the Beatriu
de Pinós postdoctoral program under the Ministry of Research and Universities of the Government of Catalonia (2022 BP 00225)
and grants PID2022-136224NB-C21, PID2022-
136224NB-C22 and 2021-SGR-872. 
FA and CP acknowledge support by the grant PID2022-
138963NB-I00  funded by
MCIN/AEI/10.13039/501100011033 and by “ERDF A
way of making Europe''.
PF is partially supported by  the STFC Consolidated Grant
ST/X000931/1.

\appendix

\section{Primitive variables recovery}\label{con2prim_appendix}

The computation of the fluxes and sources after every time step requires the knowledge of the updated values of the conserved and primitive variables.  However, not all variables are directly updated during the evolution; in particular, primitive variables that do not have their own evolution equations will have to be reconstructed from the values of the evolved variables.  This is achieved by providing a suitable conservative-to-primitive conversion (\verb|contoprim|) scheme.

In the  BDNK formulation of relativistic viscous hydrodynamics, the variables that are obtained from the evolution equations are the conserved variables $\mathbf{q}=(\sqrt{\gamma}\,E,\,\sqrt{\gamma}\,S_i)$, from the balance laws, see eqs. \eqref{eq:eq_E}--\eqref{eq:eq_Si}, and the primitive variables $\mathbf{p}_0=(\epsilon,\,v_i)$, from the first order reduction in time, see eqs. \eqref{eq:evolution_epsilon}--\eqref{eq:evolution_v}. The remaining primitive variables, $\mathbf{p}_1=(\hat\epsilon,\,\hat{\bar v}_i)$, are not directly determined by the evolution equations, and must be recovered using the \verb|contoprim|. In order to proceed, we first note that the BDNK conserved variables can be written as a linear function of the primitive variables $\mathbf{p}_1$.\footnote{This is because the viscous stress-energy tensor \eqref{eq:vis_stress_tensor} is linear in the time derivatives of the thermodynamic variables.}  In particular, we can express it in matrix form as follows:
\begin{equation}
\left(
\begin{array}{c}
E \\
S_i
\end{array}
\right) = 
\left(
\begin{array}{cc}
\mathcal{A}_{0}^{\phantom{0}0} & \mathcal{A}_{0}^{\phantom{0}j} \\
\mathcal{A}_{i}^{\phantom{i}0} & \mathcal{A}_{i}^{\phantom{i}j}
\end{array}
\right)
\left(
\begin{array}{c}
\hat\epsilon \\
\hat{\bar v}_j
\end{array}
\right) +
\left(
\begin{array}{c}
c_0 \\
c_i
\end{array}
\right)\,,
\label{eq:con2prim_gen}
\end{equation}
where the entries of the matrix $\mathcal{A}$ are given by
\begin{widetext}
\begin{align}
\mathcal{A}_{0}^{\phantom{0}0} =&~ W \bigl[- \tau_\epsilon W^2 + (\tau_p + 2 \tau_Q p'(\epsilon)) (1 -  W^2)\bigr]\,,\\
\mathcal{A}_{0}^{\phantom{0}j} = &~\tfrac{1}{3}\, v^{j} \,W^3 \Bigl[(3 \zeta + 4 \eta) (-1 + W^2) + 3 \bigl(\tau_p -  (\tau_p + 2 \tau_Q + \tau_\epsilon) W^2\bigr) (\epsilon + p)\Bigr]\,,\\
\mathcal{A}_i^{\phantom{i}0} =&~-v_{i}\, W\, \bigl[(\tau_p + \tau_\epsilon) W^2 - \tau_Q\, p'(\epsilon) (1 - 2 W^2)\bigr]\,,\\
\mathcal{A}_i^{\phantom{i}j} =&~-\delta_i^{\phantom{i}j} W \big[\eta (1-  W^2) + \tau_Q W^2 (\epsilon + p)\big]- v_{i} \,v^j\, W^3 \Bigl[\eta\bigl( 1-  \tfrac{4}{3}\, W^2\bigr) + W^2 \bigl(- \zeta + (\tau_p + 2 \tau_Q + \tau_\epsilon) (\epsilon + p)\bigr)\Bigr]\,,
\end{align}
\end{widetext}
and 
\begin{widetext}
\begin{align}
c_{0} = &-p (1 - W^2) + W^2 \epsilon +W\big(\tau_{\epsilon}\,W^2-(1-W^{2})\tau_{p}\big) \bigl[ (\epsilon + p) (-K+a^{i} v_{i} + D_{i}v^{i} + W^2\,v^{i} v^{j}  D_{i}v_{j}) + v^{i} D_{i}\epsilon\bigr] \nonumber\\
&+2\,\tau_{Q}\, W^3 \Bigl[ (\epsilon + p)\bigl( a^i v_i+ v^{i}v^{j}(- K_{ij} + W^2\, D_{i}v_{j})\bigr) + p'(\epsilon) \,v^{i}D_{i}\epsilon\Bigr]\nonumber\\
&+\tfrac{2}{3} \,\eta\,W \Bigl[(1 - W^2)(K+ 2 \,a^{i} v_{i} -  D_{i}v^{i}) + W^2 \Bigl( v^{i} v^{j} \bigl(3\, K_{ij} -  (1 + 2 W^2) D_{i}v_{j}\bigr)\Bigr)\Bigr] \nonumber\\
&+\zeta\, W (1 - W^2) (- K+ a^{i} v_{i} + D_{i}v^{i} + W^2\,v^{i} v^{j}  D_{i}v_{j})\,,\\
c_i=&~ v_{i}\, W^2 (p + \epsilon) +v_{i} (\tau_\epsilon+\tau_p) W^3 \bigl[ (\epsilon + p) (-K+a^{i} v_{i} + D_{i}v^{i} + W^2\,v^{i} v^{j}  D_{i}v_{j}) + v^{i} D_{i}\epsilon\bigr]\nonumber\\
&+\tau_Q\,\Big\{
 p'(\epsilon)W\, D_i\epsilon\nonumber\\
&\hspace{1.1cm}+W^3 \Bigl[(\epsilon + p) \Bigl(a_i + v_{i}\,a^{j} v_{j} + v^{j} \bigl(- K_{ij} + D_{j}v_{i} -  v_{i} v^{l} (K_{jl} - 2 W^2 D_{l}v_{j})\bigr)\Bigr) + 2\, p'(\epsilon) v_{i} v^{j} D_{j}\epsilon\Bigr] 
\Big\}\nonumber\\
&+\eta\,\Big\{
a_{i} W (1 - W^2) +  K_{ij} \,v^{j} W (1 + W^2) \nonumber\\
&\hspace{.9cm}- \tfrac{1}{3} W^3 \big[ v_i \big(2\, K+ a^{j}  v_{j} - 3 K_{jl}  v^{j} v^{l}- 2  D_{j}v^{j} + 4\, W^2\, v^{j} v^{l}D_{j}v_{l}\big) + 3 \,v^{j}( D_{i}v_{j} +  D_{j}v^{i}) \big]
\Big\}\nonumber\\
&-\zeta\,v_{i}\, W^3 \big(- K + a^{j} v_{j} + D_{j}v^{j} + W^2\,v^{j} v^{l}  D_{j}v_{l}\big)\,.
\end{align}
\end{widetext}
Defining a new vector as:
\begin{equation}
\left(
\begin{array}{c}
b_0 \\
b_i
\end{array}
\right)    
     = \left(
\begin{array}{c}
E \\
S_i
\end{array}
\right)-
\left(
\begin{array}{c}
c_0 \\
c_i
\end{array}
\right)\,,
\end{equation}
the \verb|contoprim| then amounts to solve the linear system 
\begin{equation}
 \left(
\begin{array}{cc}
\mathcal{A}_{0}^{\phantom{0}0} & \mathcal{A}_{0}^{\phantom{0}i} \\
\mathcal{A}_{i}^{\phantom{i}0} & \mathcal{A}_{i}^{\phantom{i}j}
\end{array}
\right)
\left(
\begin{array}{c}
\hat\epsilon \\
\hat{\bar v}_i
\end{array}
\right) =
\left(
\begin{array}{c}
b_0 \\
b_i
\end{array}
\right)\,.
\label{eq:con2prim_3p1}
\end{equation}
It is worth noting that the \verb|contoprim| procedure differs between the BDNK and perfect fluid cases. The former involves solving a linear system, i.e., inverting the matrix in \eqref{eq:con2prim_3p1}, which can be done analytically. In contrast, the perfect fluid case requires solving a nonlinear equation at every time step.

For the special case of spherical symmetry that we consider in the main text, the variables that are obtained from the numerical evolution are the the conserved variables $\mathbf{q}=(\tilde{\gamma}\,E,\tilde{\gamma}\,S_{r})$ and the primitive variables $\mathbf{p}_0=(\epsilon,\tilde{v}^r).$ Note that the physically relevant velocities $v^r, \partial_r v^r$ can be directly recovered using the definition of $\tilde{v}^r= \tfrac{1}{r}v^{r}$. The remaining primitive variables $\mathbf{p}_1=(\hat{\epsilon},\hat{\bar v}^r)$ are obtained with the \verb|contoprim|. In this case, the linear problem is given by
\begin{equation}
 \left(
\begin{array}{cc}
\mathcal{A}_{0}^{\phantom{0}0} & \mathcal{A}_{0}^{\phantom{0}1} \\
\mathcal{A}_{1}^{\phantom{0}0} & \mathcal{A}_{1}^{\phantom{0}1}
\end{array}
\right)
\left(
\begin{array}{c}
\hat\epsilon \\
\hat{\bar v}^r
\end{array}
\right) =
\left(
\begin{array}{c}
b_0 \\
b_i
\end{array}
\right)\,,
\end{equation}
with
\begin{widetext}
\begin{eqnarray*}
    \mathcal{A}_{0}^{\phantom{0}0} &=& -\frac{2 g_{rr} (v^r)^2 \tau_Q \partial_\epsilon p+\tau_\epsilon \left(g_{rr}
   (v^r)^2 \partial_\epsilon p+1\right)}{\left(1-g_{rr} (v^r)^2\right)^{3/2}}~,\\
    \mathcal{A}_{0}^{\phantom{0}1} &=& -\frac{g_{rr} v^r \left(-4 g_{rr} (v^r)^2 \eta+3 g_{rr} (v^r)^2 \left((p+\epsilon) \tau_\epsilon \partial_\epsilon p-\zeta\right)+3 (p+\epsilon) (2
   \tau_Q+\tau_\epsilon)\right)}{3 \left(1-g_{rr} (v^r)^2\right)^{5/2}}~,\\
   \mathcal{A}_{1}^{\phantom{0}0} &=& -\frac{g_{rr} v^r \left(\left(g_{rr} (v^r)^2+1\right) \tau_Q \partial_\epsilon p+\tau_\epsilon \left(\partial_\epsilon p+1\right)\right)}{\left(1-g_{rr} (v^r)^2\right)^{3/2}}~,\\
   \mathcal{A}_{1}^{\phantom{0}1} &=& -\frac{g_{rr} \left(-4 g_{rr} (v^r)^2 \eta +3 g_{rr} (v^r)^2 \left((p+\epsilon
   ) \left(\tau_\epsilon \left(\partial_\epsilon p+1\right)+\tau_Q\right)-\zeta \right)+3
   (p+\epsilon) \tau_Q\right)}{3 \left(1-g_{rr} (v^r)^2\right)^{5/2}}~.
\end{eqnarray*}
\end{widetext}

The precise expressions for the components of the  vector $(b_0,b_i)$ are cumbersome and do not provide interesting insights into the problem.  We will therefore not include them, and simply comment that they depend on both the geometry and the matter fields.

\section{Convergence tests} \label{covergencesection}

\begin{figure}
    \centering
    \includegraphics[width=1\columnwidth]{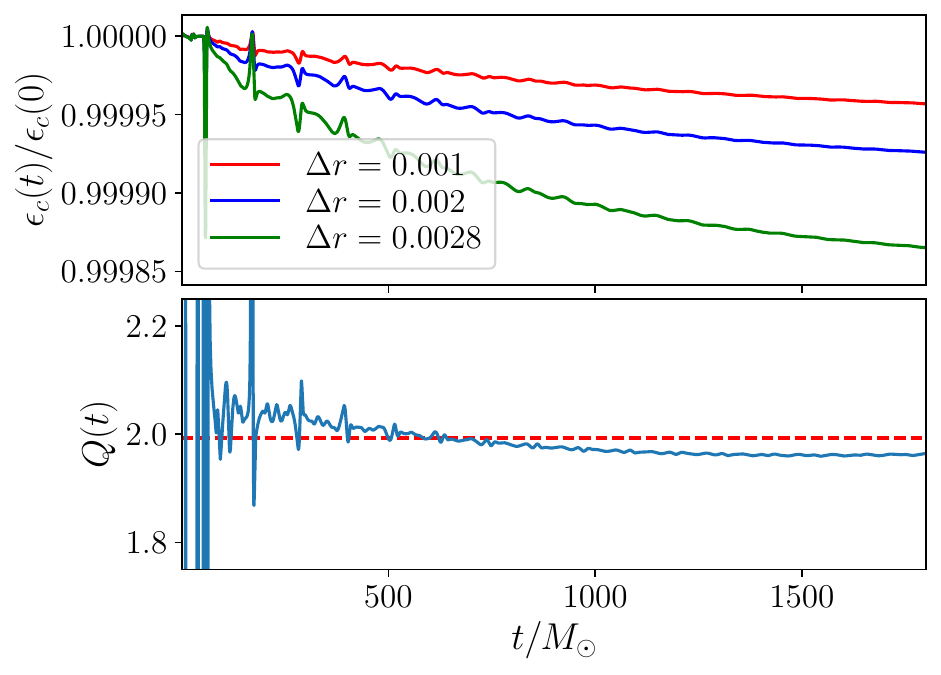}
    \caption{\emph{Convergence assessment.} \textbf{Top panel:} Convergence of central energy density evolutions is shown. This panel qualitatively suggests that increasing the resolution leads to convergence. \textbf{Bottom panel:} The convergence factor is displayed. It is obtained for resolutions that are not constant multiples of each other ($\Delta r=[0.0028,0.002,0.001] M_\odot$) using the data from \texttt{smallSB-F2}.  The red horizontal line shows the theoretically expected value of $Q$.  We find that the convergence factor achieves the theoretically expected value.}
    \label{fig:conv_dr2821}
\end{figure}

To check that the errors in our simulations are decreasing at the expected rate, we perform a point-wise convergence test on the central energy density as a function of time for one of our representative BDNK cases.  To take into account the fact that the resolutions are not necessarily constant multiples of each other, we use a more general equation to estimate the convergence factor \cite{Alcubierre}:\\
\begin{equation}
    Q=\frac{(\Delta r_l)^n-(\Delta r_m)^n}{(\Delta r_m)^n-(\Delta r_h)^n}
\end{equation}
where $\Delta r_l,\Delta r_m,\Delta r_h$ are the grid spacing of the low, middle, high resolutions respectively, and $n$ is the convergence order.  Since the data are not necessarily extracted at the same time for all resolutions (e.g. under the same CFL condition, the resolutions $\Delta r=0.002 M_\odot$ and $\Delta r=0.0028 M_\odot$ do not produce time steps that are integral multiples of each other), we use a Cubic Spline interpolator to ensure the comparison across resolutions is made under the same time step.  The Cubic Spline interpolator has a convergence order of 4, therefore is sufficient for our third order numerical scheme.\\
\\
We find that convergence is obtained at high enough resolutions.  For the resolutions $\Delta r=0.0028,0.002,0.001 M_\odot$ (see Fig. \ref{fig:conv_dr2821}), the numerical convergence asymptotes to the expected value after a short transient behaviour.  We also show in Table \ref{QNM_freq_convergence} that the QNM frequencies are reasonably stable across different resolutions.\\

\begin{table}[!h] 
\begin{center}
\begin{tabular}{||c | c c c||} 
 \hline
 $\Delta r$ & F & H1 & H2 \\ [0.5ex] 
 \hline\hline
 0.0028 & 2.69 & 4.60 & 6.36 \\ 
 \hline
 0.002 & 2.69 & 4.60 & 6.36 \\
 \hline
 0.001 & 2.67 & 4.61 & 6.33 \\
 \hline
\end{tabular}
\end{center}
\caption{\emph{Convergence of the $f$-mode frequencies in kHz}  This table shows the QNM frequencies (in kHz) of \texttt{smallSB-F2} as a function of resolution, showing that they are mostly stable against resolution change.}
\label{QNM_freq_convergence}
\end{table}
\bibliography{fluids}
\bibliographystyle{unsrt}

\end{document}